\documentclass[aps,prd,twocolumn,superscriptaddress,floatfix,longbibliography]{revtex4-2}

\usepackage{textcomp,gensymb}
\usepackage{hyperref}
\usepackage{graphicx}
\usepackage{amsfonts,amsmath,amssymb,bm,bbm}
\usepackage{color,soul}
\usepackage{slashed}
\usepackage[toc,page]{appendix}
\usepackage{flushend}
\usepackage{physics}
\usepackage{graphicx}
\usepackage{dcolumn}
\usepackage{float}
\usepackage{aas_macros}

\graphicspath{{./figures/}}

\hypersetup{
    pdfnewwindow=true,      
    colorlinks=true,       
    linkcolor=blue,          
    citecolor=blue,        
    filecolor=blue,      
    urlcolor=blue        
}

\begin{document}

\title{New Constraints on Macroscopic Dark Matter Using Radar Meteor Detectors}

\author{Pawan Dhakal}
\email{pawandhakal14@gmail.com}
\affiliation{Center for Cosmology and AstroParticle Physics (CCAPP), The Ohio State University, Columbus, OH 43210, USA}
\affiliation{Department of Physics, The Ohio State University, Columbus, OH 43210, USA}
\affiliation{Vision Dolpo, Kathmandu 44600, Nepal}
\affiliation{Open Learning Exchange Nepal, Kathmandu 44600, Nepal}

\author{Steven Prohira}
\email{prohira@ku.edu}
\affiliation{Center for Cosmology and AstroParticle Physics (CCAPP), The Ohio State University, Columbus, OH 43210, USA}
\affiliation{Department of Physics, The Ohio State University, Columbus, OH 43210, USA}
\affiliation{Department of Physics and Astronomy, University of Kansas, Lawrence, KS 66045, USA}

\author{Christopher V. Cappiello}
\email{cvc1@queensu.ca}
\affiliation{Center for Cosmology and AstroParticle Physics (CCAPP), The Ohio State University, Columbus, OH 43210, USA}
\affiliation{Department of Physics, The Ohio State University, Columbus, OH 43210, USA}
\affiliation{Department of Physics, Engineering Physics, and Astronomy, Queen's University, Kingston, Ontario, K7N 3N6, Canada}
\affiliation{Arthur B. McDonald Canadian Astroparticle Physics Research Institute, Kingston ON K7L 3N6, Canada}
\affiliation{Perimeter Institute for Theoretical Physics, Waterloo, Ontario, N2L 2Y5, Canada}

\author{John F. Beacom}
\email{beacom.7@osu.edu}
\affiliation{Center for Cosmology and AstroParticle Physics (CCAPP), The Ohio State University, Columbus, OH 43210, USA}
\affiliation{Department of Physics, The Ohio State University, Columbus, OH 43210, USA}
\affiliation{Department of Astronomy, The Ohio State University, Columbus, OH 43210, USA} 
\author{Scott Palo}
\email{scott.palo@colorado.edu}
\affiliation{Department of Aerospace Engineering Sciences, University of Colorado Boulder, Boulder, CO 80309, USA}

\author{John Marino}
\email{john.marino@colorado.edu}
\affiliation{Department of Aerospace Engineering Sciences, University of Colorado Boulder, Boulder, CO 80309, USA}

\date{April 13, 2023}

\begin{abstract}
We show that dark-matter candidates with large masses and large nuclear interaction cross sections are detectable with terrestrial radar systems.  We develop our results in close comparison to successful radar searches for tiny meteoroids, aggregates of ordinary matter.  The path of a meteoroid (or suitable dark-matter particle) through the atmosphere produces ionization deposits that reflect incident radio waves.  We calculate the equivalent radar echoing area or `radar cross section' for dark matter.  By comparing the expected number of dark-matter-induced echoes with observations, we set new limits in the plane of dark-matter mass and cross section, complementary to pre-existing cosmological limits.  Our results are valuable because (A) they open a new detection technique for which the reach can be greatly improved and (B) in case of a detection, the radar technique provides differential sensitivity to the mass and cross section, unlike cosmological probes.
\end{abstract}

\maketitle


\section{Introduction}

The particle nature of dark matter (DM) remains elusive, even after decades of ever-more sensitive astrophysical, cosmological, collider, and direct detection searches~\cite{Bertone_2005, peter2012dark, Buckley_2018, Bertone_2018, Arbey:2021gdg}. Many of these probes are based on searches for DM-nucleus elastic scattering, with null results leading to constraints in the plane of DM mass ($m_\chi$) and DM-nucleon cross section ($\sigma_{\chi N}$). For spin-independent scattering, the DM-nucleus cross section $\sigma_{\chi A}$ is usually related to the DM-nucleon cross section $\sigma_{\chi N}$ by the simple scaling relation $\sigma_{\chi A} = (\mu_{\chi A}/\mu_{\chi N})^2 A^2 \sigma_{\chi N}$, where $\mu$ denotes the reduced mass (see, e.g., Refs.~\cite{Lewin:1995rx, Schumann:2019eaa}).  However, as shown in Ref.~\cite{barnPaper}, which calculated where model independence ends, this scaling starts breaking down above $\sigma_{\chi N} \approx 10^{-31} \text{ cm}^2$, with cross sections above $\sigma_{\chi N} \gtrsim 10^{-25} \text{ cm}^2$ only possible for composite (non-pointlike) DM or for pointlike DM where the DM-nucleon interaction is attractive and has resonances (see, e.g., Refs.~\cite{Bai:2009cd, Xu:2020qjk}).

\begin{figure}[t]
    \centering
    \includegraphics[width=\columnwidth]{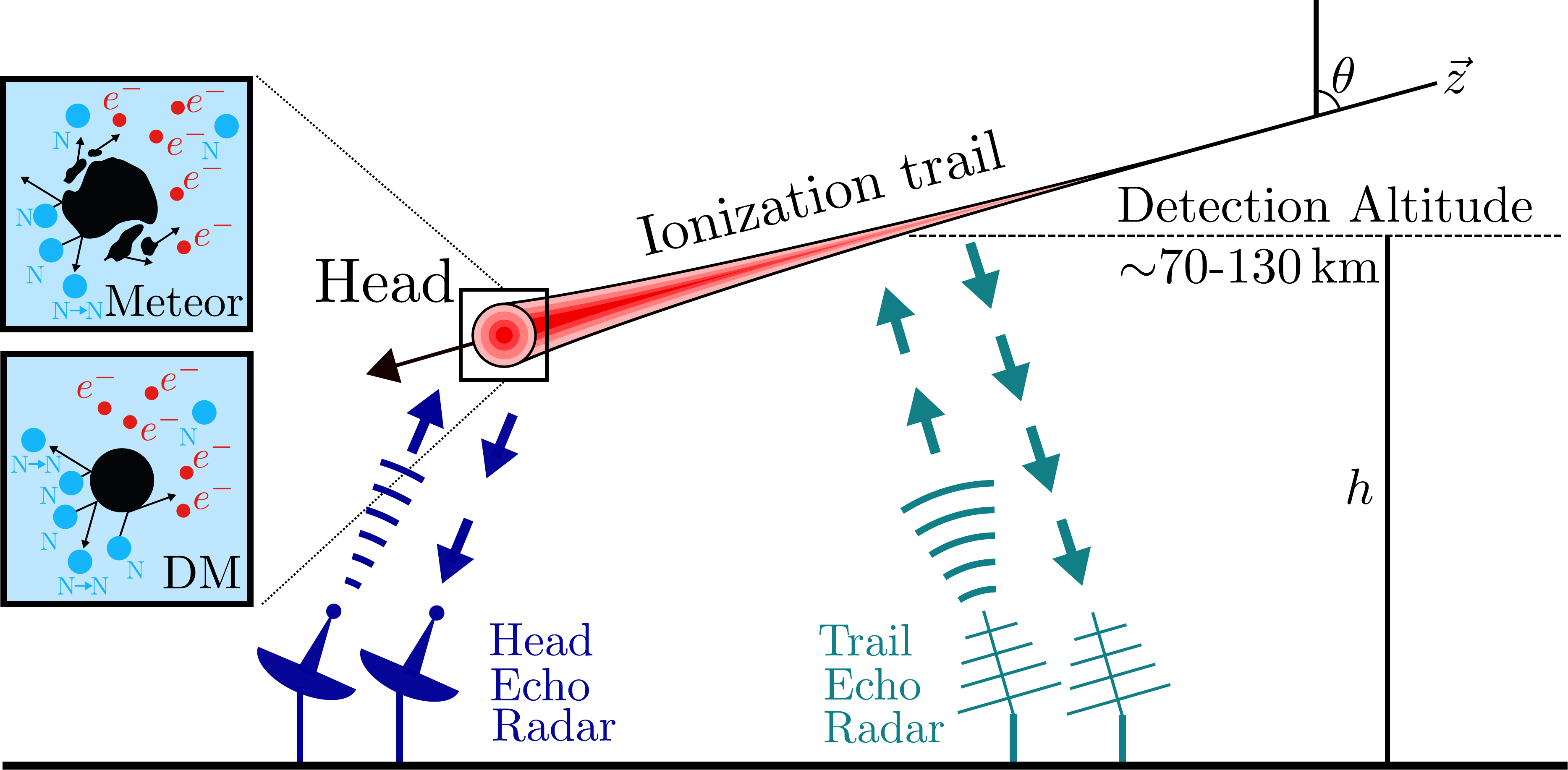}
    \caption{Radar detection of atmospheric ionization from meteors (and DM) for a simplified antenna setup. Here $\theta$ is the zenith angle, $h$ the altitude, and $\vec{z}$ the incoming trajectory.  Head-echo radar systems detect the roughly spherical ionization front around the moving object, while trail-echo radar systems detect its ionization trail. The red gradient shows a Gaussian electron density profile, which determines the type of  scattering (underdense if radio waves penetrate the ionization profile, and overdense otherwise). See text for details.
    }
    \vspace{-0.3cm}
    \label{fig:meteor_physics_cartoon}
\end{figure}

In this paper, we assume that the full DM mass density is due to macroscopic DM, for which the candidates have Planck-scale masses and non-pointlike cross sections $\sigma_{\chi N}$ equal to the geometric size of the DM, $\sigma_{\chi}$, which is vastly larger than the sizes of nuclei~\cite{barnPaper}.  A review is given in Ref.~\cite{starkmanMacroall} (see also Ref.~\cite{Carney:2022gse} for a review of ultraheavy DM in general). Such candidates are difficult to search for in direct-detection experiments, because their large masses make their fluxes very low and because their large cross sections can cause them to lose energy while passing through the overburden~\cite{Albuquerque:2003ei, Kavanagh:2017cru, bhoonah2020etching, cappiello2020new, DEAP3600_constraints, Carney:2022gse}. {\it Here we consider the atmosphere not as shielding, but as the detection volume itself --- a literal `cloud chamber.'} Some previous papers have considered the entire Earth or atmosphere as a detection volume~\cite{de1984nuclearites, porter1985search, Mack:2007xj, Gorham:2015rfa, Bramante:2019fhi, DFNconstraints}, but the detection principle presented here and the parameter space we constrain are different.  We compare the passage of macroscopic DM in the atmosphere with that of meteoroids, cosmic bodies of asteroidal or cometary origin~\cite{meteorScienceBook}.  While the energy-loss mechanisms of meteoroids and DM are different, both leave ionization deposits.   Encouragingly, the isotropic average DM flux, $\phi_\chi \sim 4 \times 10^{-4} \ (1 {\rm \ g \ } / m_\chi) {\rm \ \textrm{km}^{-2}\textrm{ hr}^{-1}}$, is always larger than the incoming meteoroid flux at a comparable mass~\cite{GRUN1985244}.

Figure~\ref{fig:meteor_physics_cartoon} shows how meteoroids are detected (and DM could be) in a simplified radar setup. Meteoroids lose energy to excitation and ionization of their surface atoms as well as to direct collisions with atmospheric nuclei.  The resulting ionization deposit acts as a conductor, which reflects radio waves.  Radio waves may be reflected from the ionization surrounding the meteoroid (a head echo) or that left behind (a trail echo).  The operating parameters for the two cases differ primarily due to the differences in radar cross sections of head and trail echoes.

We show that macroscopic DM would also produce ionization deposits detectable by meteor radar systems, here due to only direct collisions with atmospheric nuclei.  By comparing the expected numbers of head and trail echoes produced by DM with the measured meteoroid echo counts, we constrain a sizeable region in the plane of DM mass and cross section.  The concept of radar detection of cosmic-ray and neutrino-induced cascades in the atmosphere has a long history~\cite{blackett, Gorham:2000da, tara, tara_limit}, and radar-echo neutrino detection is an active area of research~\cite{t576_run2, RadarEchoTelescope:2021rca}. While other papers have considered meteor-search techniques to probe DM~\cite{DFNconstraints,DIMS:2021ipc}, this is the first to propose using atmospheric radar and to provide extensive details about the underlying physics and how to improve sensitivity.

We aim for a precision of one order of magnitude on the final results (regions of present exclusion and future sensitivity in the plane of DM mass and cross section), which is reasonable given that our final regions cover 10--15 orders of magnitude in the DM cross section.  Accordingly, we neglect a variety of uncertainties that affect our limits at the level of a factor of $\sim$2 or less.

In Sec.~\ref{sec:reviewMeteors}, we review meteor terminology and radar systems. In Sec.~\ref{sec:energyDepDM}, we calculate the rate of energy deposition into the atmosphere along the DM trajectory.  In Sec.~\ref{sec:ionizationDeposit}, we describe how the consequent ionization propagates and dissipates.  In Sec.~\ref{sec:radarDetectDMPlasma}, we calculate the equivalent radar echoing area of the resulting ionization density. In Sec.~\ref{sec:DMconstraints}, we derive constraints on macroscopic DM by comparing to observed meteor data.  In Sec.~\ref{sec:discussion}, we summarize our conclusions and discuss ways forward.


\section{Review of Radar Detection of Meteors}
\label{sec:reviewMeteors}

A \textit{meteor} is a brief burst of deposited energy in the atmosphere when a \textit{meteoroid}, moving at typical velocities of 11--70 $\textrm{km/s}$, is slowed as it moves through an increasingly dense atmosphere~\cite{meteorScienceBook}.  (For comparison, the DM average velocity is about 300 $\textrm{km/s}$.) Even at the low atmospheric densities $(\lesssim 10^{-8} \text{ g/cm}^3)$ typical of \textit{meteoric} altitudes (roughly 70--130 km, where most meteors are detected~\cite{MUpaper}), this energy loss is sufficient to vaporize and ionize surface atoms of the meteoroid (ablation) and the atmospheric gas (direct collisions), leaving ionized atoms along the meteoroid's trail.


\subsection{Meteor Terminology}

Meteors can be categorized by how bright they appear~\cite{meteorScienceBook}. Generally, the larger and heavier a meteoroid is, the more spectacular the event. If it is particularly bright, it is called a \textit{fireball}. Larger meteoroids may survive the atmosphere, reaching the Earth's surface, and are called \textit{meteorites}.  Phrases like \textit{visual meteors}, \textit{photographic meteors}, and \textit{radio/radar meteors} indicate the particular detection technique used~\cite{meteorScienceBook}.  Different techniques are most efficient for different ranges of meteoroid masses~\cite{meteorScienceBook, newInsightsRadioMeteor}.  Meteoroids of the smallest masses, like those we compare to here, are best detected with radar.  The typical mass range detected by meteor radar systems is roughly $10^{-10}$ g  to $1$ g, corresponding to a size range of roughly $10^{-7} \textrm{ cm}^2 \textrm{ to } 1 \textrm{ cm}^2$ (sporadic meteors, defined below, have densities less than $2 \textrm{ g cm}^{-3}$)~\cite{newInsightsRadioMeteor, MUpaper, meteordensities}, so these meteoroids are small as macroscopic objects but large compared to nuclei.

Meteors are also divided into classes based on their origin.  \textit{Shower} meteors are observed when Earth passes through the trails of dusty debris left by short-period comets~\cite{meteorEnvironmentLink}. Meteors from a particular shower appear to originate from a single direction called the radiant.  Sporadic meteors, or \textit{sporadics}, are produced by a diffuse, roughly isotropic background of meteoroids of cometary and asteroidal origin~\cite{meteorEnvironmentLink}. Sporadics greatly outnumber all known shower meteors and are the major contributors to the total mass influx into the atmosphere, with the flux dominated by smaller meteoroids. For the remainder of this paper, we focus on the sporadic meteors, as the DM flux in the solar neighborhood is assumed to be roughly isotropic, like the sporadic flux.

As shown in Fig.~\ref{fig:meteor_physics_cartoon}, radio waves can be reflected by different regions of ionization in the vicinity of the meteoroid body or along its trail in the atmosphere.  \textit{Head echoes} are radio reflections from the immediate, roughly spherical region of ionization surrounding the meteoroid~\cite{MUpaper}. Head echoes are strongly Doppler-shifted, as the ionized region appears to move at the meteoroid's velocity. While such ionization lasts for as long as the meteoroid is ablating, the observed duration of the head echoes is restricted by the spatial width of the radar beam. As high-power, narrow-beam radar systems are typically used to detect head echoes, the observed durations range from a few to tens of milliseconds, though meteoroids can ablate for seconds or longer~\cite{MUpaper}.  \textit{Trail echoes} (or specular meteor echoes) are radio reflections from the quasi-stationary, roughly cylindrical ionization trail left behind as a meteoroid passes through the atmosphere. These echoes may last for up to hundreds of seconds depending on the altitude, the ionization density, atmospheric diffusion rates, electron re-attachment and recombination rates, and other atmospheric effects like wind and turbulence~\cite{meteorScienceBook, forbes1995first,  forbes1999dynamics}. Trail echoes reflect at or near the transmitted frequency, as the quasi-stationary trail is subject only to wind motion, resulting in small Doppler shifts in the received signal.


\subsection{Meteor Radar Systems}

A radar system has a transmitting antenna that broadcasts a radio signal into a volume, and a receiving antenna that monitors for echoes. In monostatic radar systems, a single antenna may perform both roles or the two antennas may be located in the same area. In bistatic radar systems, the two antennas may be spatially separated by tens to thousands of kilometers, enabling the radar system to determine the meteor orbit. Novel multistatic systems with multiple antennas for both the transmitter and the receiver are also being developed~\cite{multistaticRadar}. We focus on monostatic radar systems.

When a meteoroid passes through the active volume, radio waves from a particular radar system may reflect off the head or trail ionization deposits (if large enough), depending on the geometry of the meteor velocity vector and the wave vector (k-vector) of the radio waves~\cite{meteorScienceBook}.  If the two vectors are aligned (when the meteoroid is directly approaching), radio waves will reflect off only the head ionization. For other orientations, radio waves will also reflect off the trail ionization. Monostatic radar systems are best able to detect trail echoes when the meteoroid velocity is nearly perpendicular to the wave vector (specular reflection). For such trail-echo systems, the zenith angle of the meteor and the elevation angle of the radar signal with respect to the receiver (the complement of the zenith angle) are the same. For head-echo systems, no such specular assumption is made because of the spherical shape of the radar target, and therefore reflections can be obtained at a fixed observation zenith angle for varying arrival zenith angles. Also, for meteoroids of the same size and mass, head echoes are typically weaker than trail echoes when measured by the same radar system, with the power difference being a few tens of decibels (see Ref.~\cite{meteorScienceBook}, page 33).

Radar systems typically do not detect both head and trail echoes. Since head echoes are fainter than trail echoes, the transmitting systems for head-echo radar typically have high peak transmission power and a narrow-beam antenna pattern. Modern high-power, large-aperture radar systems, like the 1\,MW (monostatic) Shigaraki Middle and Upper Atmosphere radar (SMUR) in Japan, are able to detect very faint head echoes throughout the meteoric region in the atmosphere~\cite{MUpaper}.

On the other hand, trail-echo radar systems, like the University of Colorado Boulder's 32 kW (monostatic) Antarctic Meteor (CUAM) radar, have relatively lower transmitting power but a wide-beam, all-sky illumination pattern~\cite{ AMradarlink}. These systems are cheaper to run, so the sky can be monitored for meteor trails continuously. Due to the lower operating frequencies, these systems are also more appropriate for measurement of thermospheric winds (the CUAM radar system was built for this purpose)~\cite{forbes1995first, forbes1999dynamics}. The receiving antennas of the CUAM radar system also have interferometric capabilities that allow determination of the angle-of-arrival to a few degrees~\cite{AMradarlink}. Because of the lower transmitting power, they probe a different range of meteoroid masses than head-echo radar systems.


\section{DM Energy Deposition in the Atmosphere}
\label{sec:energyDepDM}

DM with a large cross section will lose energy via elastic scattering with atmospheric nuclei as it passes through the atmosphere. We focus on the case $m_\chi \gg m_A$, for which many collisions are required to appreciably slow the DM and the DM trajectory is nearly straight. The struck nuclei recoil at velocities comparable to the DM velocity, subsequently losing energy by ionizing atmospheric atoms. We derive the DM energy loss rate as a function of the DM mass and cross section, as well as initial velocity, zenith angle, and altitude. 


\subsection{DM Scattering with Atmospheric Nuclei}

We consider elastic scattering of a DM particle with mass $m_\chi$ and initial velocity $v_{\chi i} \approx 300$ km/s with a nucleus (typically nitrogen or oxygen) in the atmosphere, with mass $m_A$ and velocity $v_{A i} \approx 0$ in comparison to $v_\chi$.  After a collision, the DM final velocity $v_{\chi f}$ is
\begin{equation}
    v_{\chi f} = v_{\chi i} \sqrt{1 - 2 \frac{m_A m_\chi}{(m_A + m_\chi)^2} (1 - \cos{\theta_{CM}})},
\end{equation}
where $\theta_{CM}$ is the recoil angle in the center of momentum frame, while all other quantities are expressed in the lab frame~\cite{BookLandauMechanics}.  For $m_\chi \gg m_A$, the DM final velocity is
\begin{align}
    v_{\chi f} & \simeq v_{\chi i} \sqrt{1 - 2 \frac{m_A }{ m_\chi} (1 - \cos{\theta_{CM}})}.
\end{align}

After each collision, the DM velocity is reduced by a negligible amount, so that many collisions are required for the DM to deposit its energy.  We average over the scattering-angle distribution by setting $\cos{\theta_{CM}} = 0$~\cite{Mack:2007xj}. Therefore, after $N$ successive collisions with atmospheric nuclei, the DM final velocity is reduced to
\begin{align}
    v_{\chi f} & \simeq v_{\chi i} \left(1 - 2 \frac{m_A }{ m_\chi}\right)^{N/2}\\
    & \simeq v_{\chi i} \exp\left(\frac{N}{2} \ln\left(1- 2 \frac{m_A}{m_\chi}\right)\right)\\
    & \simeq v_{\chi i} \exp\left(- N \frac{m_A}{m_\chi}\right)\,.
\end{align}
This equation allows us to model the evolution of the DM velocity in the atmosphere. Rearranging, we can also get the maximum number of collisions that will leave the DM above some final velocity $v_{\chi f}$:
\begin{align}
    N \simeq \frac{m_\chi}{m_A} \ln(\frac{v_{\chi i}}{v_{\chi f}}).
\end{align}
Since $v_{\chi i}$ appears inside the logarithm, the number of scattering events is insensitive to even large changes in the assumed initial velocity, and hence also to the assumed DM initial velocity distribution.

There is a maximum scattering angle in the lab frame for elastic collisions between two particles when one is initially at rest~\cite{BookLandauMechanics}. For a DM collision with an atmospheric nucleus at rest (with $m_\chi \gg m_A$), the maximum DM scattering angle is
\begin{align}
    \sin{\theta^{\textrm{max}}_{\textrm{lab}}} = \frac{m_A}{m_\chi}\,,
\end{align}
making the DM trajectory nearly straight.


\subsection{DM Velocity Evolution in the Atmosphere}

The next step is to relate $N$ to the path length in the atmosphere, $L$, and the cross section.  At constant density, 
\begin{align}
    N = \frac{L}{\lambda} = n_A \sigma_\chi L = \frac{\rho_A}{m_A} \sigma_\chi L, 
\end{align}
where $n_A$ is the atmospheric number density, $\rho_A$ is the atmospheric mass density, $m_A$ is the average nuclear mass, and $\sigma_\chi$ is the DM scattering cross section. The velocity of a DM particle evolves as 
\begin{align}
    v_{\chi f} & \simeq v_{\chi i} \exp\left(- \frac{\sigma_\chi}{m_\chi}\rho_A L \right).
\end{align}
To take into account the varying density of the atmosphere, we use the isothermal atmospheric model (see Appendix~\ref{App:AtmModels} for more details) for the mass density of the atmosphere, $\rho_A$, as a function of altitude, $h$:
\begin{equation}\label{eq:isothermalatm}
    \rho_A(h) = \rho_0 e^{-h/H},
\end{equation}
where $\rho_0 \simeq 1.3 \times 10^{-3} \textrm{ g cm}^{-3}$ is the density at sea level, and $H \simeq 7 \textrm{ km}$ is the scale height assuming an atmosphere composed of approximately $80\%$ nitrogen and $20\%$ oxygen (the relative composition of the atmosphere at meteoric altitudes and below is fairly constant).

When the density variation is taken into account, the quantity $L\rho_A$ is replaced by the integrated mass column density $X(h)$.  For objects traveling downward to an altitude $h$, the mass column density for the isothermal model can be expressed simply as
\begin{equation}
    X(h) = \int_h^{\infty} \dd{h'} \rho_0 e^{-h'/H} = X_0 e^{-h/H},
    \label{eq:atmColDensity}
\end{equation}
where $X_0 = \rho_0 H \simeq 910\; \text{g} \text{ cm}^{-2}$ is the approximate vertical column density at sea level.

For simplicity, we restrict ourselves to DM particles entering at zenith angles $0 \le \theta \le 60\degree$.  This corresponds to a solid angle of $\Omega = \pi$ sr, half the sky above the horizon.  When an object passes through the atmosphere at an angle $\theta$, the column density increases approximately by a factor of $\sec{\theta}$ (ignoring Earth's curvature):
\begin{equation}
    X(h,\theta) \simeq \frac{X(h)}{\cos{\theta}} = X_0 \sec{\theta} e^{-h/H}.
\end{equation}
%

\begin{figure}[t]
    \centering
    \includegraphics[width=\columnwidth]{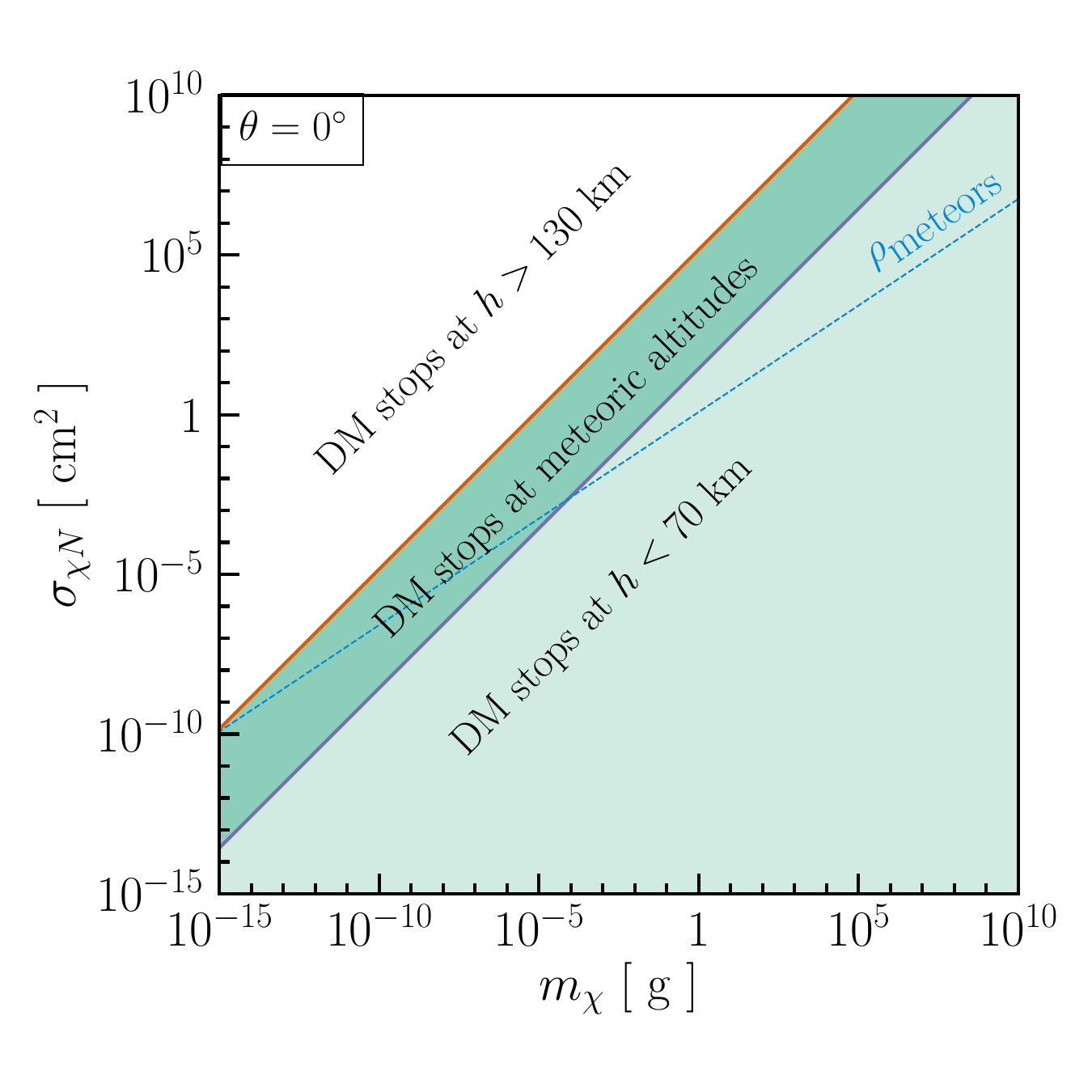}
    \caption{DM parameter space in $m_\chi$ and $\sigma_{\chi N} \; (= \sigma_{\chi})$ where DM at $\theta = 0\degree$ ``stops" (more accurately, undergoes 90\% energy loss) at altitudes where current meteor searches are sensitive (70--130 km).  Above the 130 km line (white region), DM stops before reaching meteoric altitudes.  Radar systems can detect DM that stops within the range 70--130 km (dark band) or that stops at lower altitudes but deposits enough energy in the range 70--130 km. The dashed blue line corresponds to meteors with an average density of $\sim 1 \textrm{ g cm}^{-3}$. 
    }
    \label{fig:DM_at_meteoric_alt}
\end{figure}

The final velocity of such a DM particle as a function of altitude $h$ is then
\begin{align}
    v_{\chi f}(h,\theta) & \simeq v_{\chi i} \exp\left(- \frac{\sigma_\chi}{m_\chi} X(h,\theta)\right) \\
    & \simeq v_{\chi i} \exp\left(- \frac{\sigma_\chi}{m_\chi} X_0 e^{-h/H} \sec{\theta}\right),
    \label{eq:final_velocity}
\end{align}
and the kinetic energy at a particular altitude is
\begin{align}
    E_{\chi f}(h,\theta)
    & \simeq \frac{1}{2} m_\chi v^2_{\chi i} \exp\left(- 2 \frac{\sigma_\chi}{m_\chi} X_0 e^{-h/H} \sec{\theta}\right).
\end{align}
We note that this formula, derived for collisions with individual particles, was shown in Ref.~\cite{Ellis:2021ztw} to be equivalent to the result in the fluid regime. In the limit of large DM mass, this also agrees with the continuous energy loss formalism discussed in Ref.~\cite{Emken:2018run}.

Figure~\ref{fig:DM_at_meteoric_alt} shows the region in the DM mass and cross-section plane where DM deposits most of its energy within 70--130 km, the altitude range where most meteors are detected. For large values of the reduced cross section ($\sigma_\chi/m_\chi$), DM loses nearly all of its energy high in the atmosphere. The reduced cross section at which DM loses 90\% of its initial energy above an altitude $h$ is
\begin{equation}
    \frac{\sigma_\chi}{m_\chi} = \frac{\ln{10}}{2 X_0} e^{h/H} \cos{\theta} 
    \simeq 1.3 \times 10^{-3} e^{h/H} \cos{\theta}\, \frac{\textrm{cm}^2}{g}.
\end{equation}


\subsection{Rate of Energy Deposition by DM}

The rate (technically per unit length, not time) at which a DM particle deposits energy into the atmosphere can be found by differentiating the expression for energy with respect to distance. Let the DM trajectory be from $z = -\infty$ to $z = 0$ (corresponding to sea level, $h=0$; see Fig.~\ref{fig:meteor_physics_cartoon} for an illustrative diagram), so that $\dd{h}/\dd{z}=-\cos{\theta}$. Then, the energy deposited in the atmosphere $(E_\textrm{atm})$ per unit DM path length is
\begin{align}\label{eq:energyRate}
    \dv{E_\textrm{atm}}{z} &= -\dv{E_\chi}{z} = \dv{E_\chi}{h} \cos{\theta}\\
    & \simeq \rho_A(h) \sigma_\chi v^2_{\chi i} \exp(-2  \frac{\sigma_\chi}{m_\chi}X(h,\theta))\\
    & \simeq \rho_A(h) \sigma_\chi v^2_{\chi i} \exp(-2  \frac{\sigma_\chi}{m_\chi} X_0 e^{-h/H} \sec{\theta}).
\end{align}

\begin{figure}[t]
    \centering
    \includegraphics[width=\columnwidth]{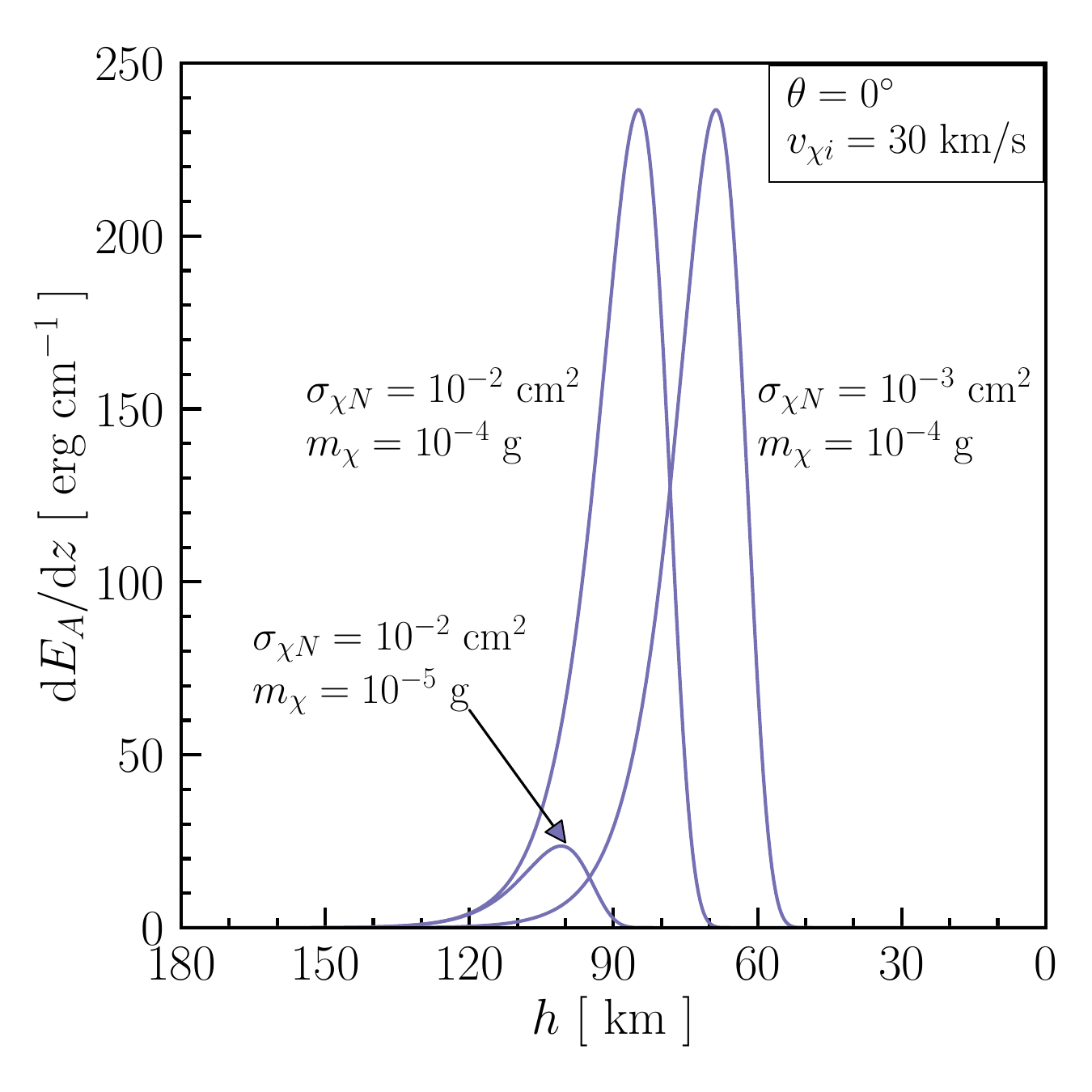}
    \caption{Rate (per length, not time) of atmospheric energy deposition by DM, which has a characteristic shape, shown by a few examples for a specific (low) DM velocity. The rate increases rapidly as DM encounters an increasing density of scatterers. When DM is slowed down enough (approximately when it reaches the altitude $h_{\textrm{peak}}$), the rate drops sharply.}
    \label{fig:DM_dEdx_ver_alt}
\end{figure}

Figure~\ref{fig:DM_dEdx_ver_alt} shows the DM energy deposition rate in the atmosphere, which has a shape reminiscent of a Bragg peak that describes the energy loss rate of charged particles in matter~\cite{ParticleDataGroup:2020ssz}, though the reasons are different.  As a DM particle passes through the atmosphere, it encounters an exponentially increasing number of scatterers in its path, increasing the energy deposition rate. Eventually, the DM is slowed enough so that the energy deposition rate peaks at an altitude $h_{\textrm{peak}}$, found by maximizing the rate with respect to the distance travelled, after which it rapidly loses energy before coming nearly to rest, 
\begin{equation}
   h_{\text{peak}} = H \ln{\left(\frac{2 X_0\, \sec{\theta}}{ (m_\chi/\sigma_\chi) }\right)}.
\end{equation}
This roughly corresponds to the altitude at which the total mass of nuclei the DM particle has scattered is comparable to the DM mass (see Ref.~\cite{2020arXiv200201476S} for an equivalent derivation of the peak altitude). For a given mass, the maximum energy deposition rate is independent of cross section (see the two peaks in Fig.~\ref{fig:DM_dEdx_ver_alt} for $m_\chi = 10^{-4}$ g),
\begin{equation}
    \dv{E_\textrm{atm}}{z}(h,\theta) \bigg|_{h=h_{peak}} = \frac{1}{2} m_\chi v^2_{\chi i} \left(\frac{1}{e H}\right) \cos{\theta},
\end{equation}
with the DM particle losing approximately $63\%$ of its initial energy before reaching $h_{\text{peak}}$ because
\begin{equation}
    \int_{h_{peak}}^\infty \dv{E_\textrm{atm}}{z}(h,\theta) dz = \frac{1}{2} m_\chi v^2_{\chi i} \left(1-\frac{1}{e}\right) \simeq 0.63  E_{\chi i}.
\end{equation}
Once DM has lost nearly all of its energy, its dynamics are controlled by thermal scattering and gravity.

The DM velocities, and hence also the nuclear recoil velocities, are $\approx 300$ km/s on average, well above the speed of sound in air. In principle, this could cause formation of a hydrodynamic shock~\cite{2020arXiv200201476S}, though we have not explored this. If a shock is formed, we expect that it would increase the ionization rate through atom-atom collisions, increasing the radar detectability.

\section{Ionization Due to DM Scattering}
\label{sec:ionizationDeposit}

After collision with a DM particle, a nucleus with mass $m_A \ll m_\chi $ and initial velocity $v_{A i} \approx 0$ recoils with velocity $v_{A f} \approx 2 v_{\chi i} \sin(\theta_{\textrm{CM}}/2) \lesssim 2 v_{\chi i}$~\cite{BookLandauMechanics}. The nucleus subsequently collides with other atmospheric nuclei, losing its energy by exciting and ionizing the surrounding gas. This results in the creation of an initial distribution of ions and electrons along the DM trajectory. The ions and electrons then move out radially together at thermal velocities of the environment ($v_A^{\textrm{th}} \sim 0.5 \textrm{ km/s}$~\cite{meteorScienceBook}) due to ambipolar diffusion (electrons initially diffuse faster than ions but an `ambipolar' electric field induced by the charge separation forces the ions and the electrons to diffuse at a common rate)~\cite{1975MNRAS.173..637J}. This dilutes the ionization density.  Other processes like electron recombination, attachment to neutral atoms and molecules, turbulence, and winds also affect the ionization density~\cite{meteorScienceBook}.


\subsection{Initial Ionization Distribution and Electron Line Density}

Following the literature on the diffusion of electrons produced along a meteor trajectory (see, e.g., Refs.~\cite{meteorScienceBook, meteoricPhenomenaBook}), we assume that the recoiling nucleus slows down to thermal velocities of the atmosphere $v_A^{\textrm{th}}$, instantaneously creating an ionization deposit with an initial radius of order the atomic atmospheric mean free path $r_0 \sim \lambda_A $. The number of ionized electrons per unit track length is called the electron line density, and is defined as

\begin{align}
    q_e \equiv 2 \pi \int_0^\infty n_{e}(r) r \dd{r},
    \label{eq:linedensity2}
\end{align}
where $n_e$ is the ionization density.

The initial ionization density follows a Gaussian profile, $n_e(r,t=0) \propto \exp(-\frac{r^2}{r_0^2})$. Plugging this into Eq.~\ref{eq:linedensity2} determines the proportionality constant in terms of $q_e$:
\begin{equation}
    n_0 = n_e(r,t=0) = \frac{q_e}{\pi r_0^2} \exp(-\frac{r^2}{r_0^2}),
    \label{eq:initialPlasmaDist}
\end{equation}
Because ions are heavier than electrons and therefore are less efficient radiators, we ignore the distribution of ions for radar detection of the ionization deposits.

The initial radius is roughly equal to the mean free path, which varies with altitude,
\begin{align}
    r_0 & \sim \lambda_A \simeq \lambda_0  e^{h/H},
\end{align}
where $\lambda_0 \simeq 10^{-6}$ cm is the atomic mean free path at sea level. At meteoric altitudes, $70$ km $\le h \le$ $130$ km, 
\begin{align}
0.02 \textrm{ cm} \lesssim r_{0} \lesssim 120 \textrm{ cm}.
\end{align}
To be conservative in our calculations of the numbers of electron-ion pairs produced by DM, we assume that the initial recoil is of a neutral atom and that all electron-ion pairs are created by subsequent atomic collisions.  We have verified with a GEANT4~\cite{agostinelli2003geant4} simulation that if a nitrogen atom at 3 keV (a typical recoil energy) is injected into the atmosphere at meteoric altitudes, nearly all of its energy is converted into ionization of the gas.  (We get very similar results with injected ions instead of atoms.)  This enables us to relate the electron line density along the DM trajectory to the energy deposition rate as
\begin{align}
    q_{e} = \frac{\dd{E_\textrm{atm}}/\dd{z}}{\expval{I}} \label{eq:lineDensity},
\end{align}
where $\expval{I}$ is the average energy required to create an electron-ion pair in the atmosphere, which depends primarily on the first ionization energies of oxygen and nitrogen atoms ($13.62$ eV and $14.53$ eV, respectively~\cite{nistIonLink}).

\begin{figure}[t]
    \centering
    \includegraphics[width=\columnwidth]{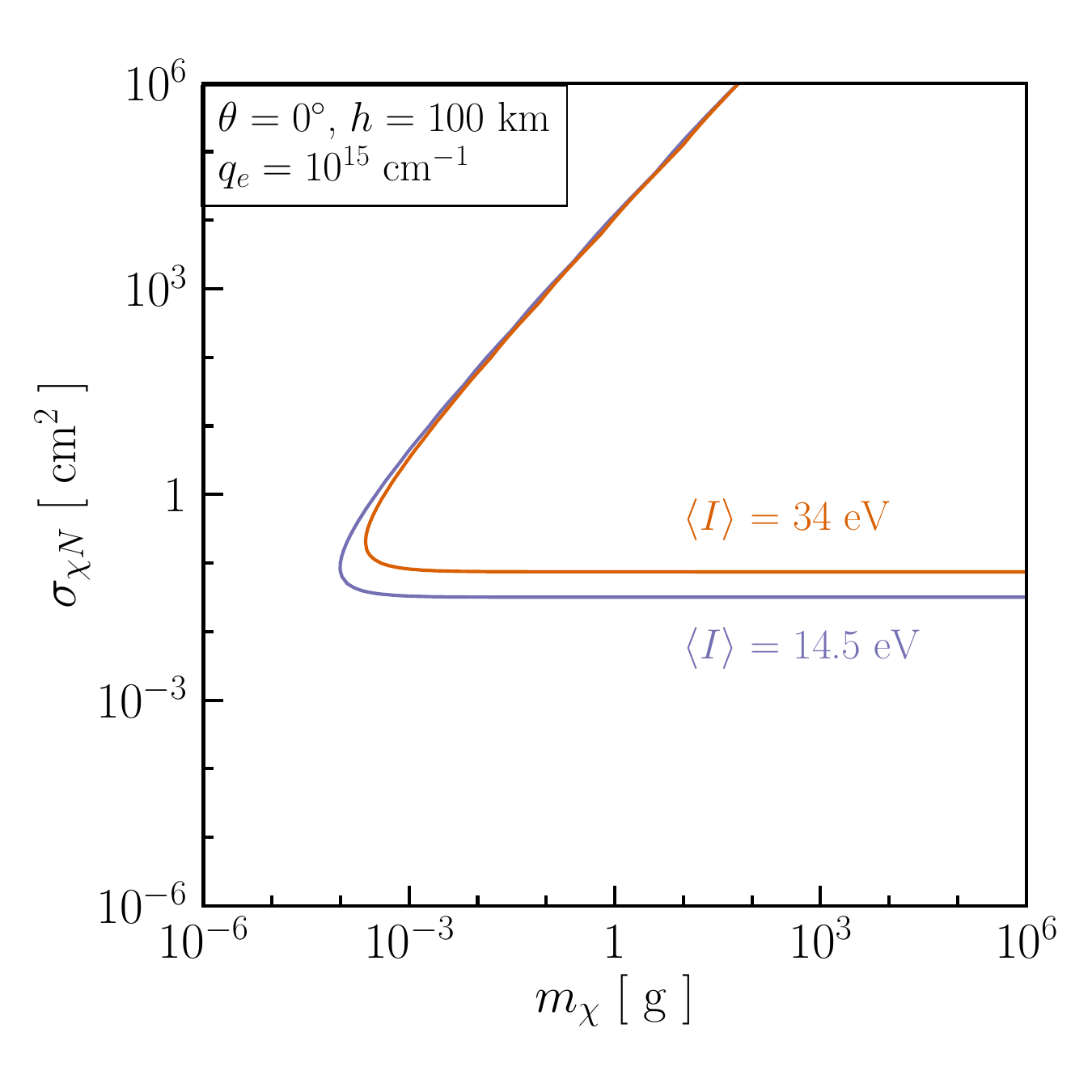}
    \caption{Contour that would produce a line density of $10^{15}$ cm$^{-1}$ for the given choices of altitude and zenith angle (above and to the right would give higher line densities).  Other combinations of electron line density, altitude, and zenith angle would produce different contours.  For different choices of the energy needed to create an electron-ion pair, the lines are nearly the same, showing insensitivity to that choice.
    }
    \label{fig:DM_mean_energy_per_ionpair}
\end{figure}

Figure~\ref{fig:DM_mean_energy_per_ionpair} shows that the range of DM mass and cross section that can produce significant ionization does not depend strongly on the choice of $\expval{I}$.  The experimentally determined value of the average energy required for a moving charged particle to create an electron-ion pair in air is $\simeq 34$ eV, i.e., 2--3 times larger than the first ionization potential of the target atom~\cite{mean_energy_ion_pair}. 

\subsection{Electron Number Density}
\label{sec:plasmaDensity}

The dominant process in the dilution of the initial ionization density is ambipolar atmospheric diffusion, during which electrons and ions in the electrically neutral gas move out and the trail expands radially at thermal velocities of the surrounding environment~\cite{meteorScienceBook}.  The initial radius expands after some time $t$ as
\begin{equation}
    r_0^2 \to r_0^2 + 4 D_a t,
\end{equation}
where $D_a$ is the atmospheric ambipolar diffusion coefficient, which increases roughly exponentially with altitude. At meteoric altitudes, the diffusion coefficient is approximately~\cite{ meteoricPhenomenaBook, meteorScienceBook} 
\begin{equation}
    D_a(h)  \simeq 0.025 \exp(0.154 \, \frac{h}{{\textrm{km}}}) \textrm{ cm}^{2}\textrm{ s}^{-1}.
\end{equation}
By solving the radial diffusion equation (equivalently, the 2D random-walk problem), 
\begin{equation}\label{eq:diffusion}
    \pdv{n_e}{t}  = \frac{D_a}{r}\,\pdv{x}{} (\grande r \pdv{n_e}{t}), 
\end{equation}
with the initial distribution given by Eq.~(\ref{eq:initialPlasmaDist}), the ionization number density is a Gaussian function of time and radial distance away from the ionization trail~\cite{meteoricPhenomenaBook, meteorScienceBook}, 
\begin{equation}
    n_{e} (r,t) = \frac{q_e}{\pi (r_0^2 + 4 D_a t)} \exp(-\frac{r^2}{r_0^2 + 4 D_a t}). \label{eqn:plasmaDensity}
\end{equation}
As above, the prefactor of $q_e$ is required to ensure that Eq.~(\ref{eq:linedensity2}) is automatically satisfied.
The total number of electrons $N_e$ along the DM trajectory is the integral over the trail length,
\begin{align}
    N_e = \int \dd{z} q_e = 2 \pi \int \dd{z} \int_0^\infty \dd{r} r n_e
\end{align}


\subsection{Electron Attachment and Ionization Lifetime}

Through observation and modeling of meteor trails, it is found that radar echoes from dense trails (with high ionization density) decay faster than predicted with only diffusion effects.  For a more realistic treatment, one must also consider electron attachment to neutral molecules and atoms, which we take into account.  We neglect some smaller corrections: electron-ion recombination (a small effect at meteoric altitudes~\cite{meteorScienceBook}), turbulent diffusion, and anomalous diffusion due to field irregularities~\cite{ meteorScienceBook, anomalousDiffusion}.

The atmospheric molecular number density is $n_A = \rho_A/m_A  \approx 2.5 \times 10^{19}\; e^{-h/H} \textrm{ cm}^{-3}$ (assuming an isothermal atmosphere composed of approximately $80\%$ nitrogen and $20\%$ oxygen) at meteoric altitudes.  This is much bigger than the ionization density $n_{e}$ characteristic of meteor echoes ($\lesssim 10^{11} \,\textrm{cm}^{-3}$, though the exact values can vary significantly based on the size and composition of the meteoroids, the entry angle, and the altitude at which the meteor appears)~\cite{meteorScienceBook}. Therefore, apart from ambipolar diffusion, electron attachment to neutral molecules and atoms dominates the dilution of the ionization density. To be sure,  we have confirmed that $n_e \ll n_A$ in all the cases we consider.

The ionization lifetime, defined as the average time taken for electrons to be captured by neutral molecules and atoms in the atmosphere, is $\tau \equiv (\beta_e n_A)^{-1}$, where $\beta_e$ is the electron attachment rate. Taking attachment into account, the line density varies with time as
\begin{align}
    q(t) = q_e e^{-t/\tau}.
\end{align}

While the ionization lifetime in the atmosphere at different altitudes is not well known and depends inversely on the ionization density, experimental data show that $ 1 \textrm{ s}  \lesssim \tau \lesssim 10 \textrm{ s}$ for altitudes $70 \text{ km} \lesssim h \lesssim 90 \text{ km}$ for an ionization density of $10^{6}$ cm$^{-3}$~\cite{plasmaLifetime}. Using $\tau = 10$ s for $h = 90$ km,
\begin{align}
    \beta_e = \frac{1}{\tau n_A } \simeq 1.5 \times 10^{-15} \textrm{ cm}^{3} \textrm{ s}^{-1}.
\end{align}
With this value of the attachment rate, $\tau \approx 0.6 \textrm{ s}$ for $h = 70$ km. Our results are not sensitive to the choice of $\beta_e$; changing it by even a factor of two has a negligible effect.  For meteoric altitudes, we then write the ionization lifetime as a function of altitude, 
\begin{align}
    \tau(h) \simeq 2.6 \times 10^{-5} e^{h/H} \textrm{ s},
\end{align}
where the exponential is large. Including this correction to Eq.~(\ref{eqn:plasmaDensity}), the ionization density can then be written as 
\begin{equation}
    n_{e} (r,t) = \frac{q_e}{\pi (r_0^2 + 4 D_a t)} \exp(-\frac{r^2}{r_0^2 + 4 D_a t} - \frac{t}{\tau}).%
    \label{eqn:plasmaDensityLifetime}
\end{equation}

\begin{figure}[t]
    \centering
    \includegraphics[width=\columnwidth]{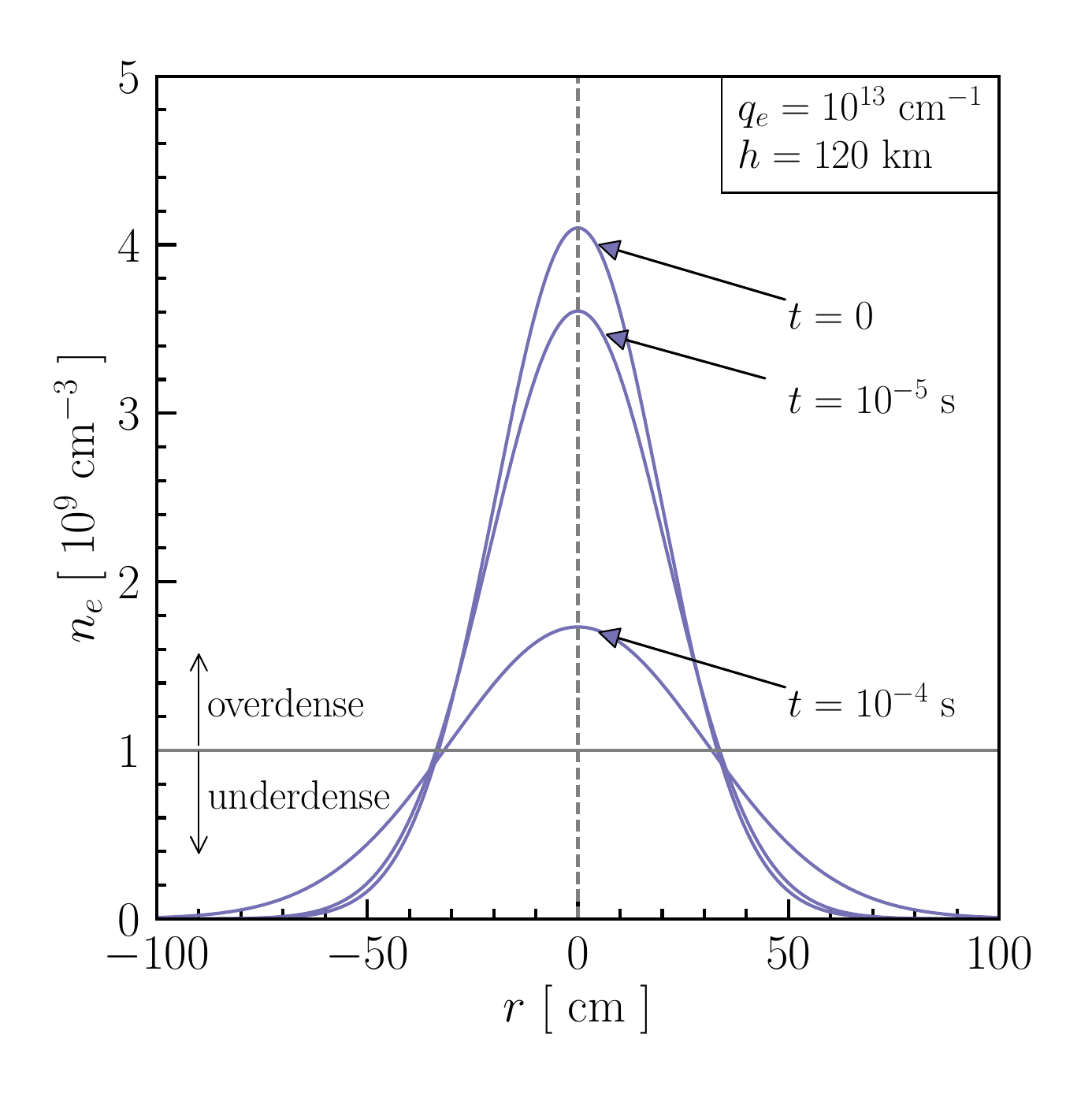}
    \caption{Electron number density as a function of radial distance from the DM trajectory for various times.  The gray horizontal line corresponds to a number density of $n_c \approx 10^9 \text{ cm}^{-3}$, which defines the boundary between underdense and overdense scattering for the SMUR and CUAM radar systems (the wavelengths for the two systems are similar). Our analysis is also sensitive to underdense ionization densities, which appear until late times.
    }
    \label{fig:plasma_number_density_vs_time}
\end{figure}

Figure~\ref{fig:plasma_number_density_vs_time} shows the ionization density as a function of radial distance from the DM trajectory for different times after the instantaneous (at $t = 0$) formation of the ionization deposit with initial radius $r_0$.  Note that both underdense echoes (echoes from ionization deposits that radio waves can penetrate) and overdense echoes (from deposits that radio waves cannot penetrate) can be detected, and the horizontal line here is simply for reference.  These two regimes are discussed in detail in the next section.

\section{Radar Detection of DM Signals}
\label{sec:radarDetectDMPlasma}

An important parameter in the analysis of radar reflections off ionization deposits is the equivalent echoing area of the radar target, called the radar cross section (RCS).  (For the relationship between RCS and optical magnitudes used in visual detection of meteors, see Ref.~\cite{ohsawa2020relationship}.)  Apart from the ionization density and the electron number density, the RCS also depends on the radar wavelength and polarization, as well as the relative geometry of the radar station and the target.  Following the treatment on radio echo theory for meteor trails in Ref.~\cite{meteorScienceBook}, we calculate the RCS for head and trail echoes from DM-induced atmospheric ionization deposits.


\subsection{Radar Cross Section}

For a monostatic radar system, the RCS of an ionized target can be inferred from the power received by the receiver using the radar equation:
\begin{equation}
    \sigma_{\text{RCS}} = \frac{(4 \pi)^3 P_r R_t^2 R_r^2}{P_t G_t G_r \lambda^2},
    \label{eq:radarEquation}
\end{equation}
where $\lambda$ is the wavelength of the interrogating radio wave.  The transmitter is characterized by its power ($P_t$), antenna gain ($G_t$), and its range (distance; $R_t$). The receiver is characterized similarly, where $P_r$ includes thermal noise and the scattered signal. Essentially, the RCS is the ratio of received power to transmitted power, once antenna effects and distance losses have been accounted for.

The term RCS, with dimensionless units of decibels per square meter (dBsm), can also refer to the logarithm of $\sigma_{\textrm{RCS}}$ relative to a square meter, so that $\sigma_{\textrm{RCS}} = 10^4 \textrm{ cm}^2$ corresponds to $0$ dBsm:
\begin{equation}
    \textrm{RCS} = 10 \log_{10}\left({\frac{\sigma_{\text{RCS}}}{10^4 \textrm{ cm}^2}}\right) \text{ dBsm}\,.
    \label{eq:RCSdBsm}
\end{equation}

We neglect polarization effects, i.e., the orientation of the incident electric field with respect to the target. For both radar systems considered here, the transmitter and receiver are circularly polarized in the same plane and are co-located (for SMUR, they are the same antenna). For a head-echo system, we neglect polarization effects because the reflector is approximately spherical, as discussed below. For a trail-echo system, we neglect them because polarization loss will occur only for the most inclined arrival directions, where the flux is minimal. 


\subsection{Underdense and Overdense Echoes}

The dielectric constant $\kappa$ of an ionization deposit with electron number density $n_e$ is~\cite{meteorScienceBook} 
\begin{equation}
    \kappa = 1 - \frac{n_e c^2 }{\pi \nu^2}r_e \simeq 1 - \left(\frac{\nu_p}{\nu}\right)^2,
\end{equation}
where $r_e \approx 2.8 \times 10^{-13} $ cm is the classical electron radius, $\nu$ is the radio frequency (36.17 MHz for the CUAM radar and 46.5 MHz for the SMUR radar), and
\begin{equation}
    \nu_p = \sqrt{\frac{c^2 r_e}{\pi} n_e}  \simeq 8956 \sqrt{\frac{n_e}{\textrm{cm}^{-3}}} \textrm{ Hz}
    \label{eq:plasmaFreq}
\end{equation}
is the plasma frequency. If $\nu > \nu_p$, then $\kappa > 0$ so that the incident radio wave penetrates the ionization deposit and electrons scatter the radio wave independently. The resulting echo is called an \textit{underdense} echo. If $\nu < \nu_p$, then $\kappa < 0$ and the radio wave does not penetrate the ionization column, as secondary collisions between electrons become important. In this case, the electrons in the ionization deposit oscillate collectively at the radio frequency and hence reflect the wave back like a metallic conductor. The resulting echo is called an \textit{overdense} echo. The transition between underdense and overdense echoes is defined by a critical ionization density, $n_c$, through Eq.~(\ref{eq:plasmaFreq}). For the radar systems we consider, $n_c \approx 10^9 \, \textrm{cm}^{-3}$ as both use similar radio wavelengths. Setting the ionization density in Eq.~(\ref{eqn:plasmaDensityLifetime}) equal to $n_c$, we get
\begin{equation}
    n_c =  \frac{q_e}{\pi (r_0^2+4 D_a t)} \exp(-\frac{r_c^2}{r_0^2 + 4 D_a t} - \frac{t}{\tau}),
    \label{eq:criticalRadiusDensity}
\end{equation}
where $r_c$ is the critical radius of the ionization column within which $\kappa \le 0$. This radius defines the boundary of the ionization column from where the radio wave is totally reflected for overdense echoes.

The two radio scattering regions (underdense and overdense) are useful because they are each described by simple analytical expressions using relevant quantities.  We emphasize that both types of echoes are detectable.  The boundary between the two regions is typically described by a transitional value $q_{\textrm{tr}}$ of the electron line density found by setting $r_c^2 = r_0^2+4D_at=\lambda^2/4\pi^2$ such that $r_c$ bounds a volume large and dense enough to attenuate the incident wave by $1/e$~\cite{meteorScienceBook}. Ignoring electron attachment effects ($\tau \to \infty$) --- which is valid for low values of $q_e$ (see Fig.~\ref{fig:plasma_radius_line_dens_attachEffect}) --- we get
\begin{align}
    q_{\textrm{tr}} = \frac{e}{4 r_e} \simeq 2.4 \times 10^{12} \textrm{ cm}^{-1}, 
\end{align}
which is independent of the incident radio frequency, and is used as a standard benchmark value for the over/under dense transition, even though in practice this transition {\it is} frequency dependent (see Ref.~\cite{meteorScienceBook}, page 215).


\subsection{Effective Ionization Radius for Overdense Echoes}

For overdense echoes, we can define an effective ionization radius $r_p$ (also called the plasma radius), taking the maximum of the critical radius $r_c$ defined above. Inverting Eq.~(\ref{eq:criticalRadiusDensity}), we get the critical radius as a function of time,
\begin{align}
    r_c^2 (t) = (r_0^2 + 4 D_a t) \left(\ln{\frac{q_e}{\pi n_c (r_0^2 + 4 D_a t)}} - \frac{t}{\tau}\right).
\end{align}

\begin{figure}[t]
    \centering
    \includegraphics[width=\columnwidth]{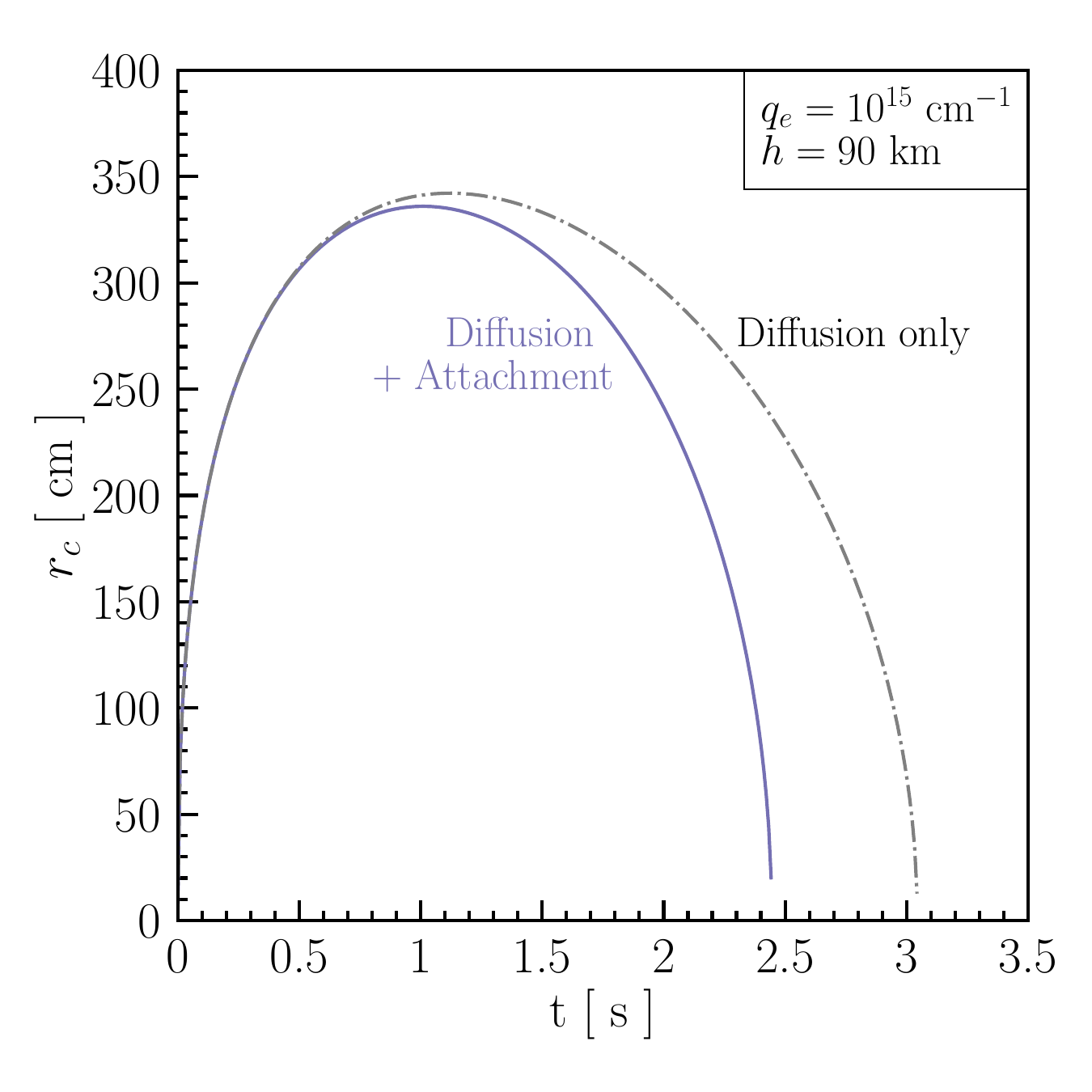}
    \caption{Critical ionization radius at which overdense scattering occurs as a function of time.  The critical radius grows to a maximum before going back to zero. The time for this defines the overdense echo duration. When electron attachment effects are included along with ambipolar diffusion, the ionization density decreases more quickly, reducing the size of the radius and the echo duration.}
    \label{fig:criticalRadius_elecAttachEffect}
\end{figure}

Figure~\ref{fig:criticalRadius_elecAttachEffect} shows that the critical radius grows to a maximum value $r_c^{\textrm{max}} \equiv r_p$ and is
\begin{eqnarray}
    r_p \equiv && \, r_c^{\textrm{max}} = \sqrt{D_a \tau W(\eta) (2+W(\eta))} \label{eqn:plasmaRadius}, \\ 
    \eta =&& \frac{q_e}{2 \pi D_a \tau n_c} \textrm{exp}\left(\frac{r_0^2}{4 D_a \tau }-1\right), \nonumber
\end{eqnarray}
where $W(\eta)$ is the Lambert-W or product-log function defined by $\eta = W(\eta) e^{W(\eta)}$.

Note that if electron attachment effects are ignored, the radius of the ionization deposit, found in a similar way using Eq.~(\ref{eqn:plasmaDensity}), equivalently by taking the limit $\tau \to \infty$ of Eq.~(\ref{eqn:plasmaRadius}) above, is simplified to 
\begin{align}
    r_p = \sqrt{\frac{q_e}{\pi e n_c}}.
\end{align}

\begin{figure}[t]
    \centering
    \includegraphics[width=\columnwidth]{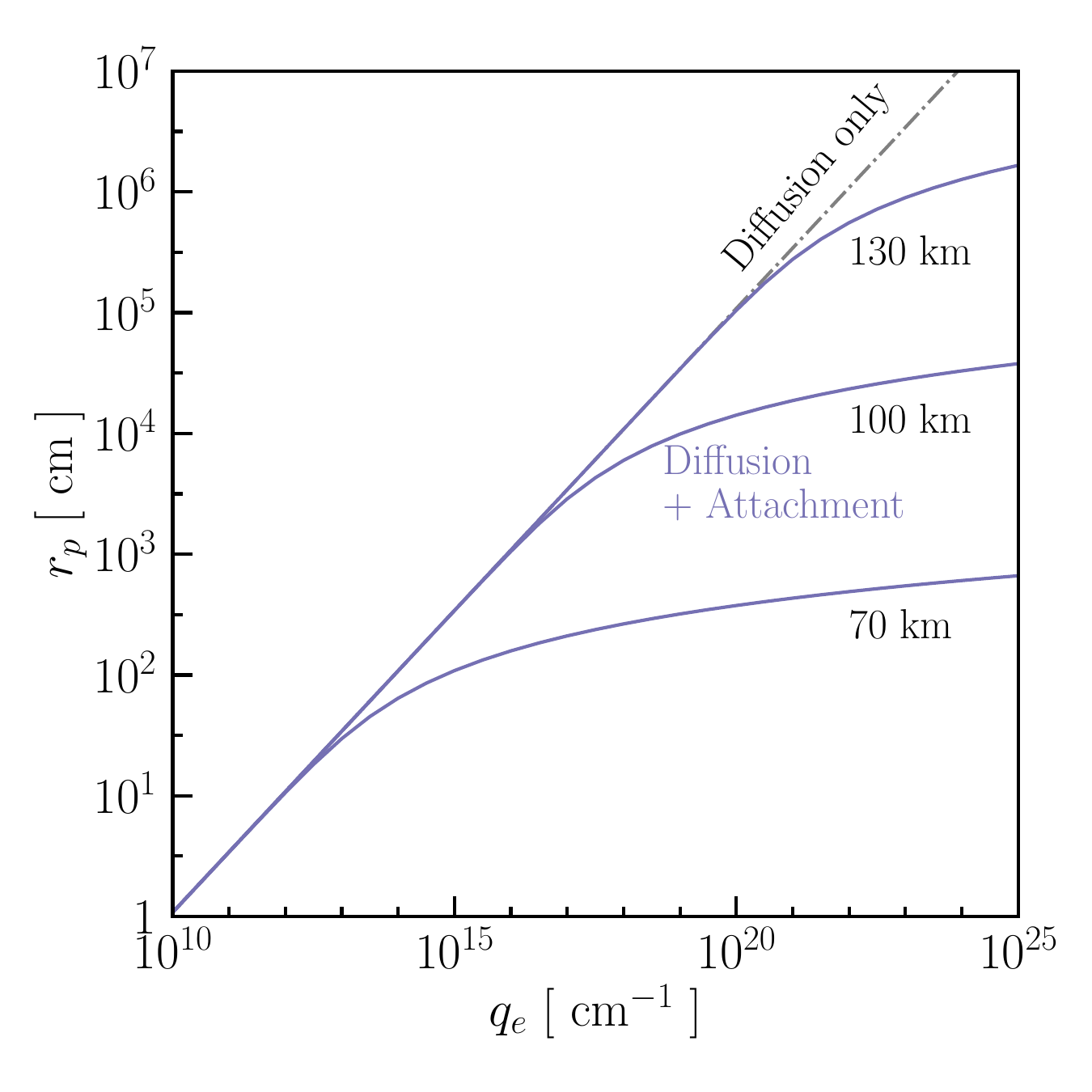}
    \caption{Radius of the ionization deposit as a function of electron line density at three altitudes, taking ambipolar diffusion into account.  When electron attachment effects are also considered, the ionization radius $r_p$ and the overdense echo duration are suppressed for larger line densities.} 
    \label{fig:plasma_radius_line_dens_attachEffect}
\end{figure}

Figure~\ref{fig:plasma_radius_line_dens_attachEffect} shows that the size of the radius is suppressed for large electron line densities when attachment effects are considered along with ambipolar diffusion.

\begin{figure}[t]
    \centering
    \includegraphics[width=\columnwidth]{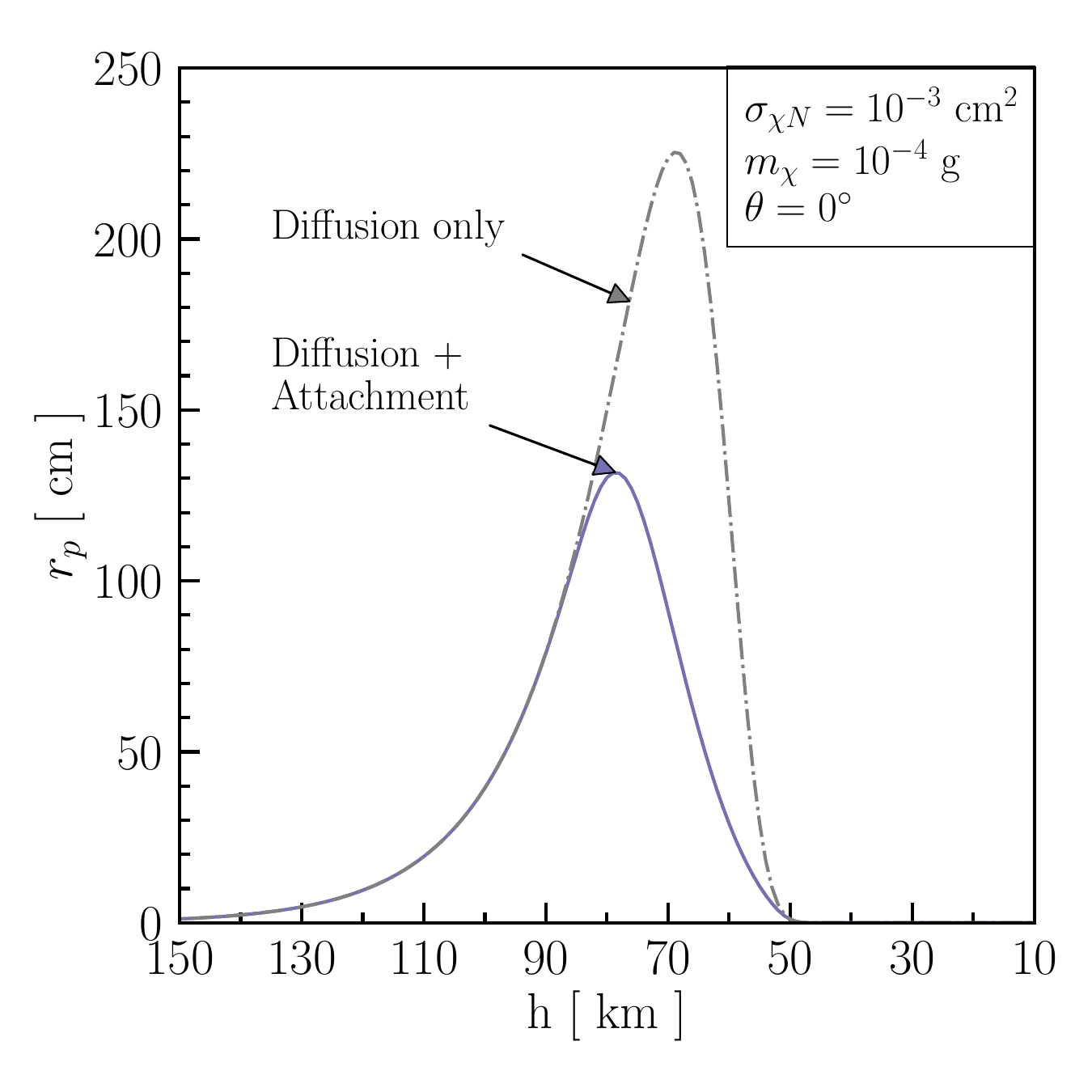}
    \caption{Radius of the ionization deposit as a function of altitude as DM traverses the atmosphere, shown for one example. When electron attachment effects are considered, the size and the profile of the radius are both modified compared to when only ambipolar diffusion is considered.}
    \label{fig:plasma_radius_comparison_diffusion_attachment}
\end{figure}

Figure~\ref{fig:plasma_radius_comparison_diffusion_attachment} compares the effective radius for overdense scattering with and without attachment as a function of altitude for a particular DM candidate. While we plot the radius for very large values of the line density to illustrate the effect of attachment clearly, it is to be noted that very large values of $q_e$ ($\gg 10^{16} \textrm{ cm}^{-1}$) are not physical~\cite{meteorScienceBook}, as the ionization density must be less than the atmospheric number density. Note that, although it is not written as a function of altitude, $q_e$ implicitly depends on altitude through its dependence on both atmospheric density and the DM velocity. This produces the altitude dependence in Fig.~\ref{fig:plasma_radius_comparison_diffusion_attachment}.

\subsection{RCS for Head Echoes}

Meteor head echoes are characterized by radio reflections off spherical ionization distributions, surrounding the meteoroid and co-moving with it~\cite{plasmaANDMassHeadEcho, plasmaDistMeteorsProfiles, simulatedHeadEcho}. Therefore, we describe the ionization deposit in the immediate neighborhood of a DM particle as a spherical Gaussian distribution in Eq.~(\ref{eqn:plasmaDensityLifetime}) centered on the particle. Although the actual shape of the ionization deposit may be more ellipsoidal (with the electron density falling off more quickly behind the parent particle), the assumption of a spherical deposit with a Gaussian profile (for the number density of electrons) is good, as demonstrated by simulations and empirical measurements~\cite{headEchoModelArecibo,meteorheadplasma_FDTDmodel}.  For initial velocities $v_{\chi i} \gtrsim 70 \textrm{ km/s}$, the deviation from spherical geometry becomes more pronounced, but this has a small effect on the radar cross section.

The head ionization is a near perfect electrically conducting sphere, with the RCS characterized by the size of the effective radius of the sphere $r_p$ relative to the radio wavelength $\lambda$~\cite{meteorheadplasma_FDTDmodel}. If $r_p \gg \lambda$ (optical scattering), the RCS approaches a constant value equal to the cross sectional area of the head plasma,
\begin{equation}
    \sigma_{\textrm{RCS}} = \pi r_p^2.
    \label{eq:RCSheadoptical}
\end{equation}
If $r_p \ll \lambda$ (Rayleigh scattering), the RCS is 
\begin{equation}
    \sigma_{\textrm{RCS}} \simeq 144 \pi^6 r_p^{6} \lambda^{-4}.
    \label{eq:RCSheadrayleigh}
\end{equation}
If $r_p \sim \lambda$ (Mie scattering), the RCS oscillates about the average value $\pi r_p^2$. For simplicity, we take the average value as the RCS for Mie scattering.  


\subsection{RCS for Trail Echoes}

In underdense trails, electrons scatter independently but coherently. The scattering cross section of a free electron for monostatic radar systems is $\sigma_e = 4 \pi r_e^2$~\cite{meteorScienceBook}. The total power received at the receiving antenna can be calculated by adding the contributions to the electric field vector from all the electrons in a line element of the trail where they scatter in phase~\cite{meteorScienceBook}. 

When the diameter of the cylindrical trail is small in comparison to the radio wavelength $\lambda$, the total quasi-instantaneous power received from all electrons from some finite section of the trail can be expressed as~\cite{meteorScienceBook} 
\begin{equation}
    P_r(t=0) \simeq \frac{P_t G_r G_t \lambda^2}{(4\pi)^3 R_0^4} \frac{R_0 \lambda \sigma_e q_e^2}{2}   \exp(-\frac{8\pi^2 r_0^2}{\lambda^2}),
    \label{eq:maxPowertrail}
\end{equation}
where $R_0$ is the minimum range along the trail, and electron attachment effects are ignored (valid for underdense trails with low ionization densities)~\cite{meteorScienceBook}. The majority of the received power comes from the first Fresnel zone around the point of closest approach to the radar station~\cite{meteorScienceBook}. The power from additional Fresnel zones alternate in phase and largely cancel out. The received power at the moment of ionization formation ($t = 0$) is scaled due to the finite initial width of the column. Taking ambipolar diffusion into account, the echo power decays as 
\begin{equation}
    P_r(t) = P_r(t=0) \exp(-\frac{32\pi^2 D_a t}{\lambda^2}).
\end{equation}
Note that as shown in Fig.~\ref{fig:plasma_radius_line_dens_attachEffect}, electron attachment effects are negligible for underdense echoes ($q_e \le q_{\textrm{tr}}$). 

The maximum power registered at the receiver is $P_r = P_r(t=0)$, which occurs when the meteor leaves the first Fresnel zone. Since the range varies slowly near $R_0$ --- the point of closest approach to the radar station --- and the length of the section of trail contributing to the majority of the received power is small compared to the range, $R \simeq R_0$~\cite{meteorScienceBook}. For a meteor entering the atmosphere at zenith angle $\theta \neq 0$ and altitude $h$, $R_0 = h/ \sin{\theta}$. Replacing Eq.~(\ref{eq:maxPowertrail}) in Eq.~(\ref{eq:radarEquation}), the maximum underdense trail-echo RCS (for backscatter) is then (for $q_e \le q_{\textrm{tr}}$),
\begin{align}
    \sigma_{\textrm{RCS}} \simeq \frac{h}{2 \sin{\theta}} \lambda \sigma_e q_e^2   \exp(-\frac{8\pi^2 r_0^2}{\lambda^2}).
    \label{eq:RCStrailUnderdense}
\end{align}

The ionization trail through the atmosphere would be perfectly cylindrical if the DM velocity were infinite (assuming the trail has already expanded to the initial radius $r_0$) and if the atmospheric density was uniform. Since DM velocity is finite but larger than the radial diffusion velocities of electrons ($v_A^{\textrm{th}} \sim 0.5 \textrm{ km/s}$~\cite{meteorScienceBook}) by 2--3 orders of magnitude (the typical DM velocity is $v_{\chi} \sim 300 \textrm{ km/s}$), the trail is more conical than cylindrical. Because both the initial radius $r_0$ and the diffusion coefficient $D_a$ exponentially decrease with altitude, the conical ionization trail but has an exponential taper~\cite{meteorScienceBook}. Note that these details are also true for ionization trails produced by meteors, which have typical velocities of 70--130 $\textrm{ km/s}$.

For overdense trails, the ionization is a near perfect electrically conducting cylinder of radius $(r_p \gg \lambda)$~\cite{meteorScienceBook}, in which case the RCS is
\begin{equation}
    \sigma_{\textrm{RCS}} = \pi R_0 r_p \simeq \frac{\pi r_p h}{\sin{\theta}},
    \label{eq:RCStrailOverdense}
\end{equation}
for incident spherical waves from a source at a perpendicular distance $R_0$~\cite{meteorScienceBook}. The power delivered by scattered waves to the receiver is then given by Eq.~(\ref{eq:radarEquation}). 


\section{DM Constraints From Radar Data}
\label{sec:DMconstraints}

With all ingredients in hand, we calculate constraints on macroscopic DM from the non-observation of excess signals in radar meteor observations. Using Eqs.~(\ref{eq:RCSheadoptical}--\ref{eq:RCSheadrayleigh}) and Eqs.~(\ref{eq:RCStrailUnderdense}--\ref{eq:RCStrailOverdense}) for the RCS and Eq.~(\ref{eq:lineDensity}) for the line density of the ionization deposit along the DM trajectory through Earth's atmosphere, we probe macroscopic DM candidates that would produce meteor-like head and trail echoes detectable by the SMUR and CUAM radar systems, respectively. The range of the RCS and of the initial entry velocity $v_{\chi i}$ that we consider depends on the experimental setup (including the radio wavelength $\lambda$) and the analysis techniques used by the radar systems.


\subsection{Calculational Approach}
\label{sec:DMconstraints_calcapproach}

The equations for the radar cross section $\sigma_{RCS}$ depend on the DM mass and cross section through the electron line density and the plasma radius, given in Eqs.~(\ref{eq:lineDensity}) and~(\ref{eqn:plasmaRadius}), respectively. Using these, we convert the DM velocity distribution for a particular DM mass and cross section into an RCS distribution. Note that both the altitude and the zenith angle must be specified to compute the RCS for both head echoes and trail echoes; while the equations for trail echoes depend explicitly on the altitude and zenith angle, those for head echoes depend indirectly on these quantities via the energy deposition rate given in Eq.~(\ref{eq:energyRate}).

For the DM velocity distribution at the top of the atmosphere $f(v_{\chi i})$, we use the Standard Halo Model as parameterized in Ref.~\cite{Evans:2018bqy}, with a velocity dispersion of $270\textrm{ km/s}$, and accounting for the motion of the Sun around the Galaxy. Meteor radars (like the SMUR and CUAM radars~\cite{MUpaper, AMradarlink}) are typically tuned to scan for meteors within the altitude range of 70--130 km (where most meteors are detected), with entry velocities of 11--70 km/s at 130 km altitude~\cite{MUexpSettings}. To account for the radar sensitivity to altitude and velocity, for each DM mass and cross section we use Eq.~(\ref{eq:final_velocity}) to calculate the maximum entry velocity at the top of the atmosphere such that the DM slows down to meteoric velocities at 130 km. This restricts the velocity distribution at the top of the atmosphere for each DM candidate (for details, see Appendix~\ref{App:VelocityDistributions}). 

For each DM mass and cross section, we then convert the restricted velocity distribution to a detected RCS (defined by the maximum RCS produced by DM within the meteoric altitude range) distribution for DM that enters the atmosphere at a particular zenith angle. We consider only incoming DM particles with zenith angles $\theta \le 60\degree$, where the curvature of the Earth can be neglected and the atmospheric mass column density in Eq.~(\ref{eq:atmColDensity}) is minimally corrected. Finally, by summing the detected RCS distributions over allowed zenith angles, we get the RCS spectrum for each DM candidate.

\begin{figure*}[t]
    \centering
    \includegraphics[height=\columnwidth]{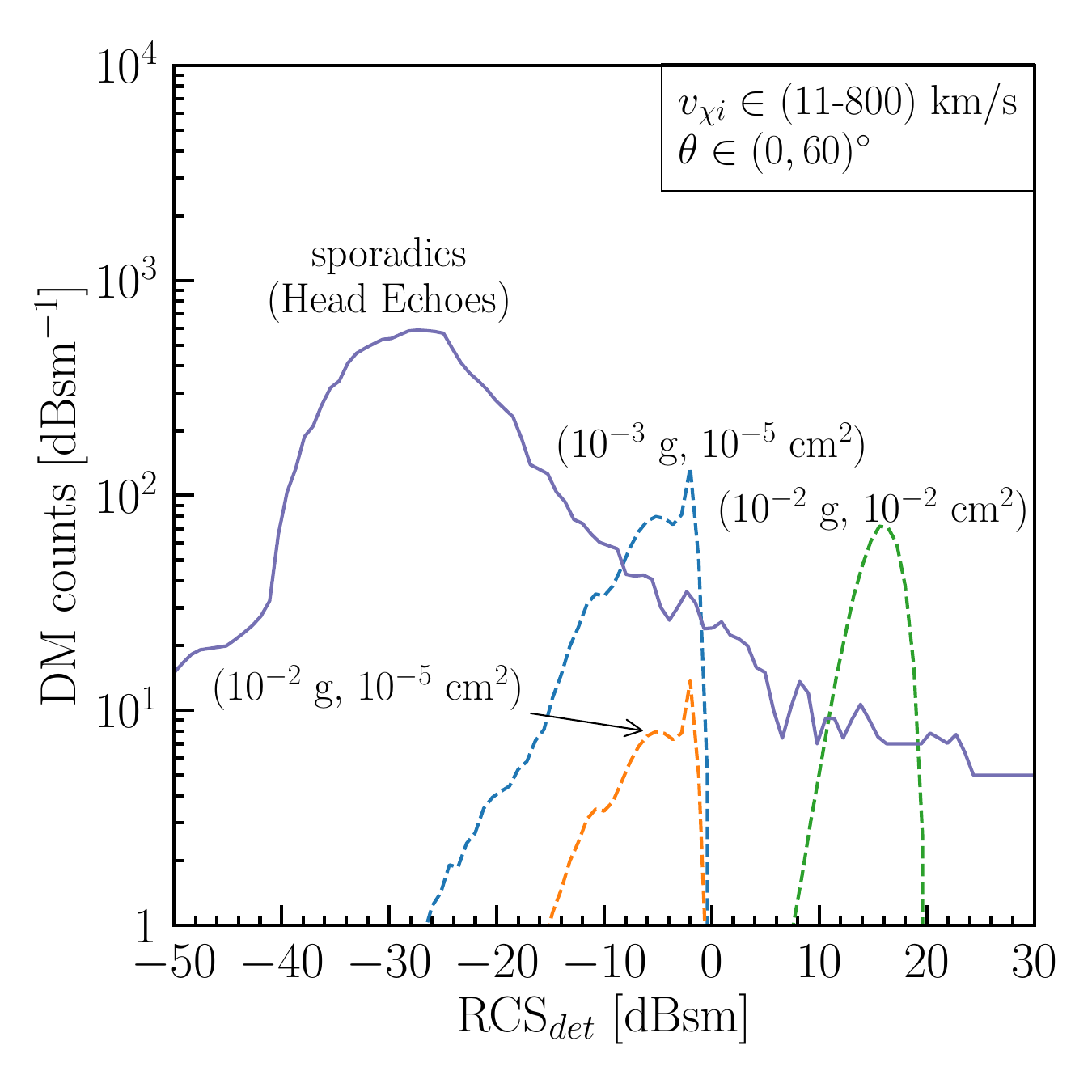}%
    \includegraphics[height=\columnwidth]{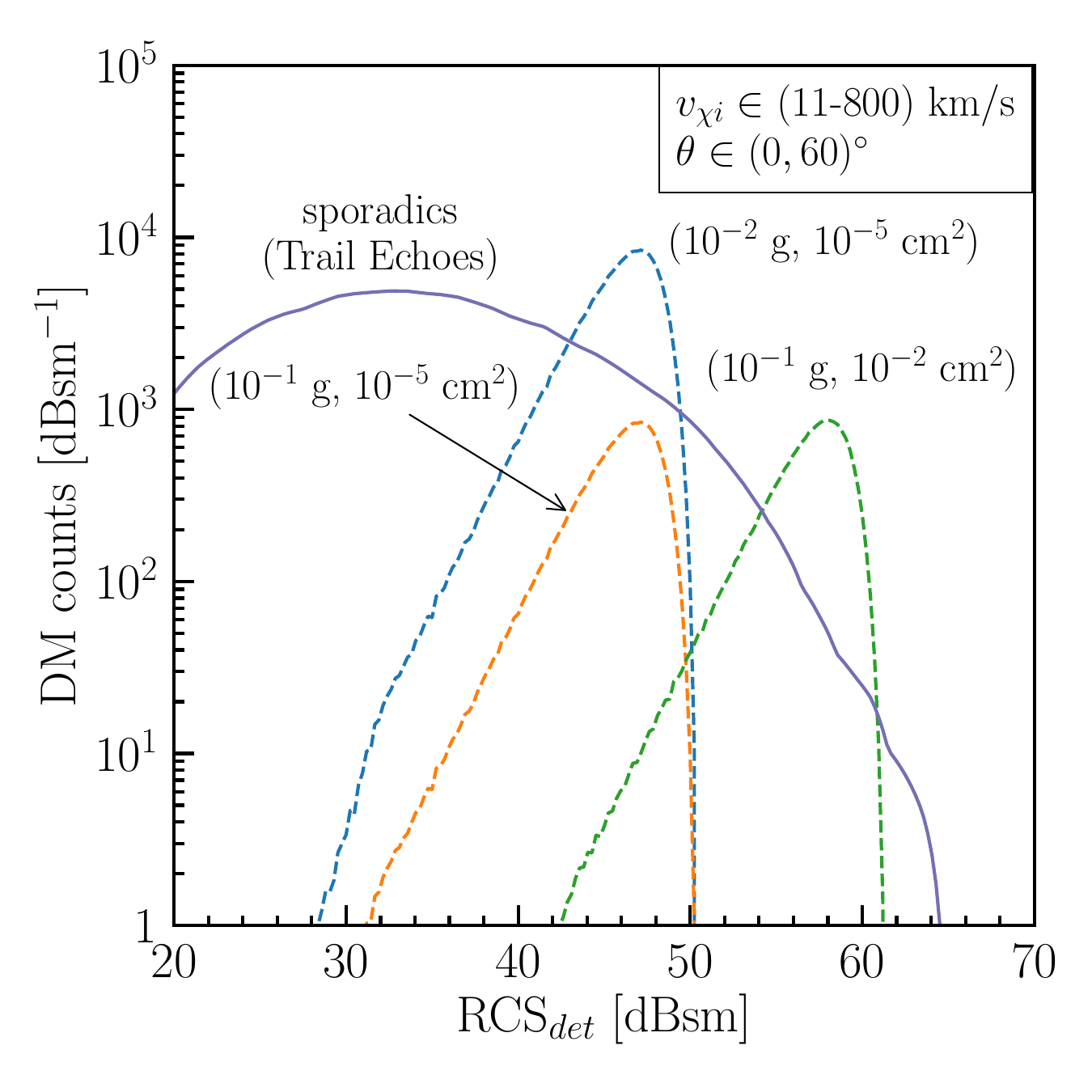}%
    \caption{Comparison of possible DM signals (dashed) to observed data  for selected masses and cross sections, using the full DM velocity range (11--800 km/s) at the top of the atmosphere and the zenith angle range $< 60\degree$. DM candidates whose counts exceed the measured data are excluded. For the same DM mass, larger cross sections correspond to larger values of the detected RCS. For the same DM cross section, larger masses correspond to fewer expected counts, due to the DM flux scaling as $1/m_\chi$. 
    {\bf Left:} Head-echo DM counts per unit RCS, shown for a few examples together with the observed head-echo data for sporadic meteors~\cite{MUpaper, MUexpSettings}.
    {\bf Right:} Trail-echo DM counts per unit RCS, shown for a few examples together with the observed trail-echo data for sporadic meteors~\cite{AMradarlink}.}
    \label{fig:head_trail_spectra}
\end{figure*}

Figure~\ref{fig:head_trail_spectra} shows example spectra for different DM masses and cross sections (integrated over zenith angles), compared to the SMUR and CUAM data~\cite{MUpaper, AMradarlink}. For head echoes detectable by the SMUR radar system $(\lambda \sim 645 \textrm{ cm})$, the RCS varies from $-50 \textrm{ to } 30 \textrm{ dBsm}$~\cite{MUpaper}. For trail echoes detectable by the CUAM radar system $(\lambda \sim 829 \textrm{ cm})$~\cite{AMradarlink}, we assume that the RCS varies from $20 \textrm{ to } 70 \textrm{ dBsm}$ based on our calculations using raw trail-echo count data from a week of observations by the CUAM radar system, which involves some uncertainties that are small relative to our precision goals.

\subsection{Data Analysis}

To set our limits, we conservatively allow that all observed echoes could be DM signals, even though they are likely all backgrounds due to meteors. We rule out a DM mass and cross section if the total number of DM events in any unit-RCS bin is significantly larger than the observed number of meteors in that bin. The rest of this subsection describes the details of setting this limit.  

For trail echoes, we restrict the data to meteors with $\theta \le 60\degree$, matching the restriction on DM mentioned in Sec.~\ref{sec:DMconstraints_calcapproach} (relaxing the restriction would increase the fluxes of both DM and meteors by a factor of $\lesssim 2$ and would have little effect on the signal-to-background ratio). For head echoes, we are unable to make this cut, as information on the zenith angle is unavailable, but we still only consider DM with $\theta \le 60\degree$ (making our results for head echoes conservative). Our analysis is thus only sensitive to DM arriving from a solid angle of $\Omega = \pi$ sr, or 1/4 of a full sphere; integrating the incoming flux over $\cos\theta \, d\cos\theta$ also to account for the component of the flux perpendicular to the surface gives 3/16 of a full sphere.  As a result, the flux of DM (per unit velocity) our analysis is sensitive to is
\begin{align}
    \dv{\phi_\chi}{v_\chi} = \frac{3}{16}\frac{\rho_\chi}{m_\chi} v_\chi f(v_\chi)\,,
    \label{eq:DMFLUXreduced}
\end{align}
where $f(v_\chi)$ is the fraction of DM particles at velocity $v_\chi$, and $\rho_\chi \approx 0.3\ \textrm{GeV}\textrm{ cm}^{-3}$ is the DM mass density at Earth's position~\cite{Bertone_2005}. For $m_\chi = 1\ \textrm{g}$ and $v_\chi = 300\ \textrm{km/s}$, $\dv*{\phi_\chi}{v_\chi} \simeq 3 \times 10^{-7}\  \textrm{km}^{-2}\ \textrm{hr}^{-1} \textrm{(km/s)}^{-1}$.

Because the energy deposition in the atmosphere depends on the DM velocity $v_{\chi}$, we can convert the velocity spectrum of Eq.~(\ref{eq:DMFLUXreduced}) into an RCS spectrum, using the equations of the previous two sections. $\phi_{\chi}$ then represents the total flux in a given RCS range. We assume that the number of detected DM particles per unit RCS follows a Poisson distribution, for which the PDF is given by
\begin{equation}
    p(n) = \frac{\mu_\chi^n e^{-\mu_\chi}}{n!},
    \label{eq:poissonDist}
\end{equation}
where $\mu_\chi = \phi_\chi \times A_{\textrm{det}} \times T_{\textrm{obs}}$ is the expected number of radar echoes produced by DM per unit RCS for a radar with effective detector area $A_{\textrm{det}}$ and total observation duration $T_{\textrm{obs}}$.  We exclude a given DM mass and cross section by conservatively requiring that the DM spectrum never be higher than the observed meteor spectrum at 95\% CL. That is, DM is ruled out if for any RCS, the CDF, $P(n \le N_m) \le 0.05$, where $N_m$ is the observed meteor count per unit RCS.

The detector area, also called the equivalent radar collection area, depends on the radar antenna gain pattern and can be expressed as a function of the RCS, $A_{\textrm{det}} = A_{\textrm{det}}(\textrm{RCS})$~\cite{MUpaper}. For the SMUR radar system, $A_{\textrm{det}}$ ranges from $1$ km$^{2}$ for RCS of $-50$ dBsm to  $10^3$ km$^{2}$ for $30$ dBsm~\cite{MUpaper}; for the head-echo data, we use $T_{\textrm{obs}} = 33$ hr~\cite{MUpaper}. For the CUAM radar system, we calculated $A_{\textrm{det}} \simeq 3\times10^{4}$ km$^{2}$ over the entire RCS range due to the all-sky nature of the antenna beam; for the trail-echo data we use, $T_{\textrm{obs}} = 118$ hr.

\begin{figure*}[t]
    \centering
    \includegraphics[height=\columnwidth]{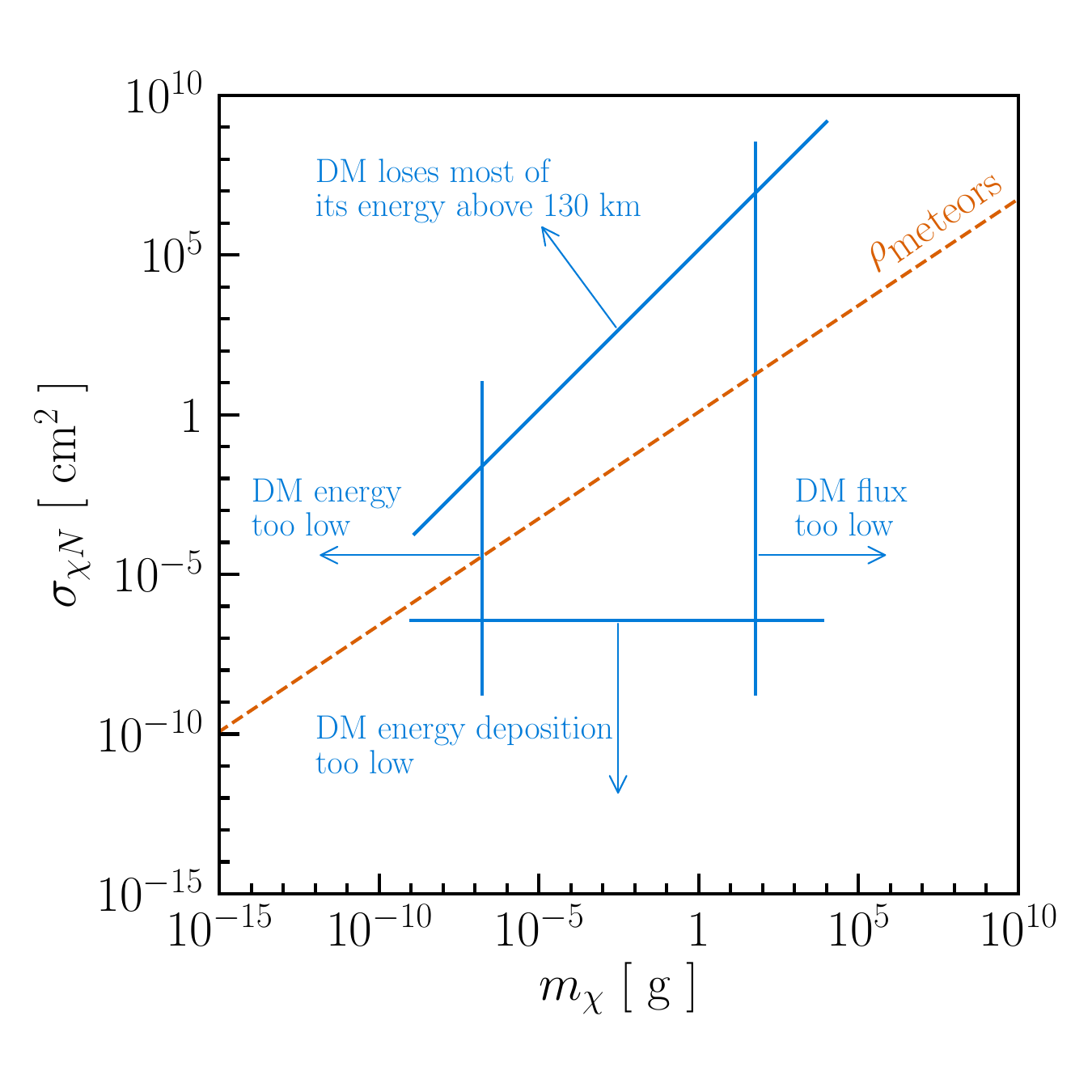}%
    \includegraphics[height=\columnwidth]{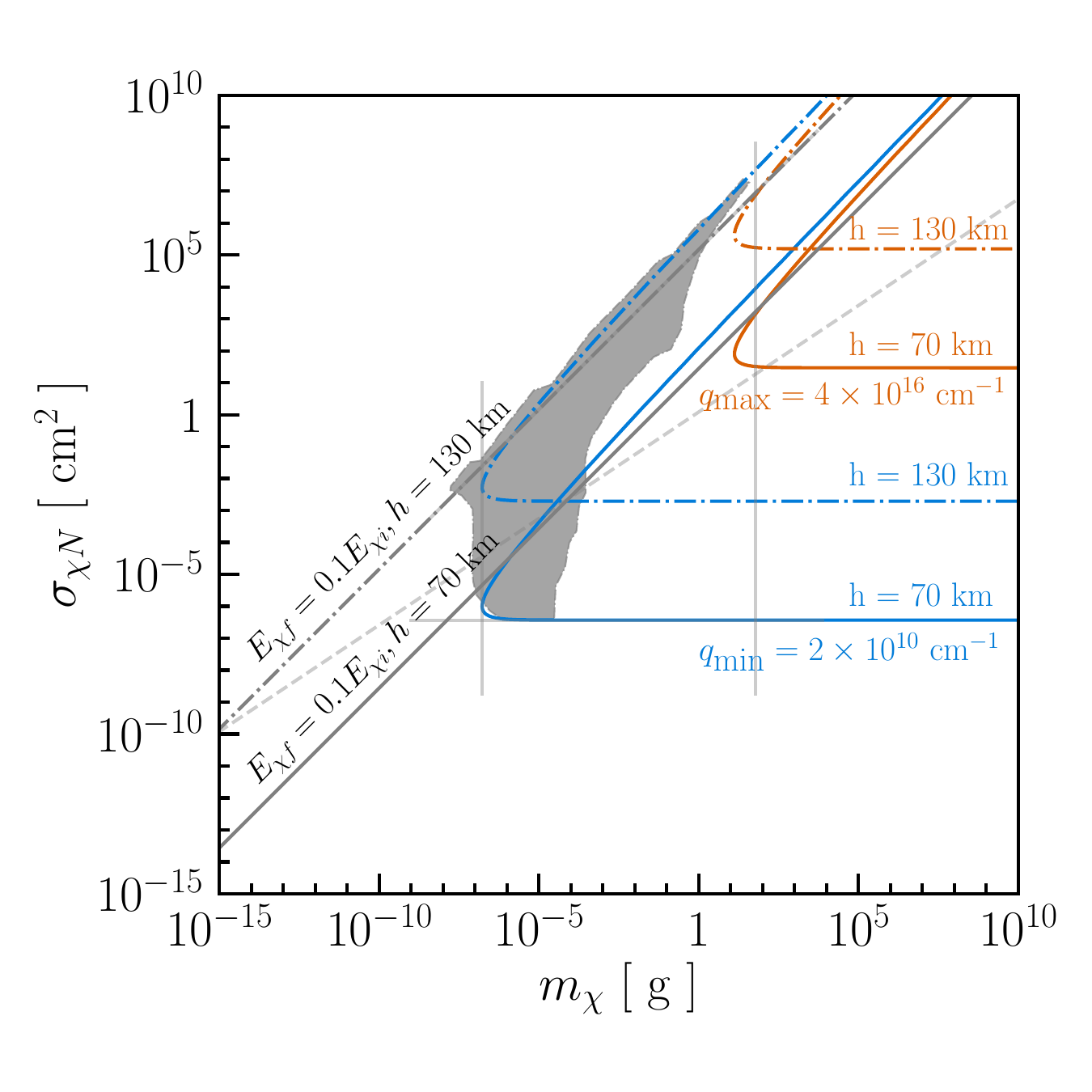}%
    \caption{Estimated sensitivity in the DM mass and cross section plane compared against the actual constraints for trail echoes detectable by the CUAM radar.  {\bf Left:} Estimates of the boundaries (blue solid), assuming zero backgrounds, for trail-echo constraints (see text for details). Also shown is a line (orange dashed) corresponding to meteors with average density of $\sim 1 \textrm{ g cm}^{-3}$.  {\bf Right:} Trail-echo constraints (gray filled region), derived by including all relevant factors, overlaid on top of the estimated region. Also shown are lines corresponding to DM candidates that lose $90\%$ of their initial energy at 130 km (gray dash-dotted) and 70 km (gray solid) altitudes, and contours corresponding to the minimum/maximum detectable line densities for the CUAM radar, or equivalently the minimum/maximum detectable RCS, (blue/orange) at 130 km (dash-dotted) and 70 km (solid) altitudes.  
    } 
    \label{fig:estimatesTrailEchoes}
\end{figure*}

Figure~\ref{fig:estimatesTrailEchoes} shows the constraints for trail echoes detectable by the CUAM radar. In the left panel, we show the estimated boundaries for the constraints using the minimum detectable RCS (lower edge), the minimum energy for DM to survive to below 130 km altitude (left edge), the largest mass that still has a large enough flux for potential detectable events (right edge), and the maximum reduced cross section $(\sigma_\chi/m_\chi)$ for which DM loses nearly all of its energy above 130 km altitude (ceiling). In the right panel, we overlay on top of the estimated boundaries the actual constraints for DM using trail-echo data, and a few descriptive lines that explain the boundaries of the exclusion region. We show the corresponding plot for head echoes detectable by the SMUR radar in App.~\ref{App:ExcRegDetails}. 
%


\subsection{Final DM Constraints}

Figure~\ref{fig:full_exclusion_region_final} shows the radar-derived exclusion regions in the plane of DM mass and cross section.  To calculate the regions, we find the detected RCS spectrum for each DM candidate as shown in Fig.~\ref{fig:head_trail_spectra} and plot the DM mass and cross section in the plane if the spectrum exceeds the meteor data. For current constraints, the velocity distribution for each candidate is restricted so that DM velocity is within the meteoric velocity range at 130 km altitude, as described in Sec.~\ref{sec:DMconstraints_calcapproach} and App.~\ref{App:VelocityDistributions}. 

For the projected sensitivities, we use the full initial-velocity range of $\sim$ 11--800 km/s, where $11\textrm{ km/s}$ is the escape velocity from Earth and $\sim 800\textrm{ km/s}$ is approximately the sum of the escape velocity from the Galaxy $(\sim 550\textrm{ km/s})$ and the velocity of the Sun $(\sim 220\textrm{ km/s})$.  For meteor velocities $v > 70\textrm{ km/s}$, we assume that $N_m \approx 0$ because of the extremely low flux of such meteors, but defer to future experiments to conclusively test this.  (More details about the meteor and DM velocity distributions are given in  Appendix~\ref{App:VelocityDistributions}.) If future experiments are sensitive to very fast-moving meteors, our limits could improve by orders of magnitude in both mass and cross section, even with the same exposure.

We also show prior DM constraints. The light gray region is excluded by observations of the Milky Way satellite population~\cite{cosmologyConstraints} and cooling of Galactic gas clouds~\cite{bhoonah2020detectinggascloud}, while the dark gray region is excluded by observations of long-lived white dwarfs (\cite{Graham:2018efk}, but see also Ref.~\cite{SinghSidhu:2019tbr}, a reanalysis that produces a substantially smaller region by using a different density profile and more conservative treatment of a thermonuclear runaway).  Relatively large cross sections have been probed by experiments sensitive to interactions in the lower atmosphere: particle detectors on a satellite, ``Skylab"~\cite{bhoonah2020etching}, and searches for optical flashes with the Desert Fireball Network, ``Fireballs"~\cite{DFNconstraints}.  Relatively small cross sections have been probed by a variety of shallow and deep underground dark-matter or repurposed experiments: ``Chicago"~\cite{cappiello2020new}, ``DAMA"~\cite{PhysRevLett.83.4918, bhoonah2020detectinggascloud}, ``DEAP-3600"~\cite{DEAP3600_constraints}, ``Ohya"~\cite{bhoonah2020etching}, and ``Mica"~\cite{starkmanMacroall}.  The ``Humans" region is constrained by null observations of unique human injuries/death by DM~\cite{humanImpacts}.  Future sensitivities (not shown) to ultraheavy DM have been derived based on collisions with stars~\cite{Das:2021drz}, signals in IceCube~\cite{Bai:2022nsv}, and tracks in quartz~\cite{Ebadi:2021cte}. Ref.~\cite{Dessert:2021wjx} also places constraints from collisions with stars, but at masses above the range we show.

Though our limits overlap with cosmological constraints, our results are complementary in several ways: they are independent of cosmological models or assumptions; they explicitly focus on composite DM, which is not typically the focus of cosmological studies; and they have the advantage of differential sensitivity to the DM mass and cross section.  In addition, the sensitivity of our approach can be significantly improved.


\section{Conclusions and Future Work}
\label{sec:discussion}

While it is usually assumed that DM interacts weakly, it remains possible that it interacts strongly but has escaped detection by being very massive, so that its number density is low~\cite{Chung:1998ua, Faraggi:1999iu, Albuquerque:2000rk, Enqvist:2001jd, Bai:2018dxf, Grabowska:2018lnd, Hong:2020est, Barman:2021ugy}.  Such macroscopic DM might not reach terrestrial detectors, instead losing a significant fraction of its energy through elastic scattering with nuclei in the overburden.

\begin{figure}[t]
    \centering
    \includegraphics[width=\columnwidth]{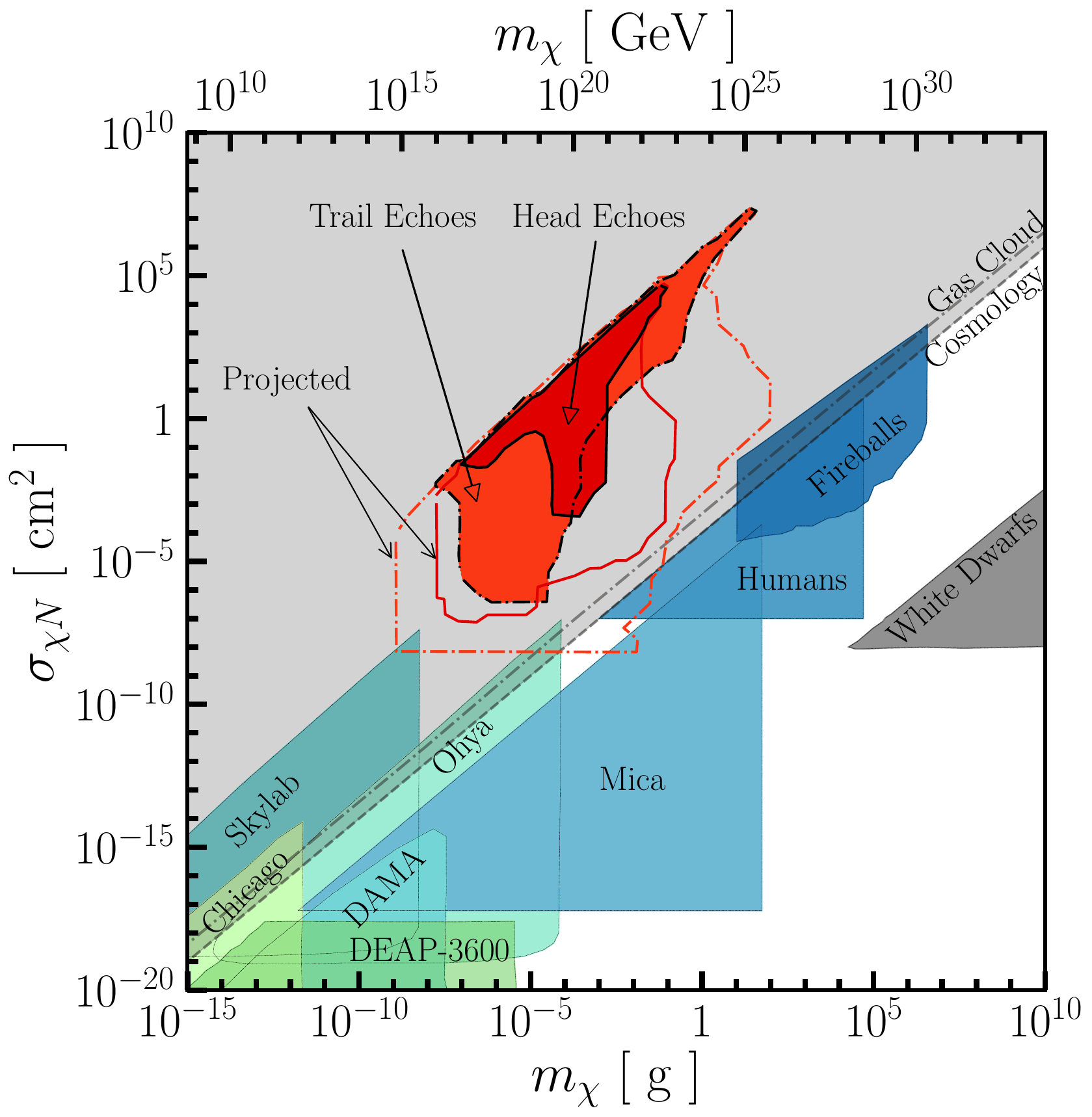}
    \caption{New meteor-radar constraints on macroscopic DM in the plane of mass ($m_\chi$) and DM-nucleon cross section ($\sigma_{\chi N} = \sigma_\chi$). Firm constraints from head echoes are shown in red (solid fill and black solid outline) and from trail echoes are shown in lighter red (solid fill and black dash-dotted outline). The transparent regions with solid red outline for head echoes and dash-dotted lighter red outline for trail echoes are projected sensitivities based on using the full DM velocity distribution. Also shown are existing constraints from astrophysics and cosmology (gray) and DM detectors (blue/green), taken from Refs.~\cite{starkmanMacroall, cosmologyConstraints, humanImpacts, DFNconstraints, cappiello2020new, bhoonah2020etching, DEAP3600_constraints, bhoonah2020detectinggascloud}.}
    \label{fig:full_exclusion_region_final}
\end{figure}

Here we consider the atmosphere as a detection volume, and radar as the probing method.  We show that macroscopic DM particles passing through the atmosphere can produce ionization deposits that are detectable with radar systems. We model the spatial evolution of the ionization over time, taking into account ambipolar diffusion and reattachment effects, to accurately determine the detectability of the resulting ionization density.  

Figure~\ref{fig:full_exclusion_region_final} shows that existing data, sensitive only to low velocities at meteoric altitudes, can be used to search for macroscopic DM, even without a dedicated analysis, constraining a wide range of parameter space.   Our constraints are model-independent in the sense that no particular model of composite DM is assumed, but we note that there are composite models, e.g., those of Refs.~\cite{Enqvist:2001jd, Grabowska:2018lnd}, that can lie in our parameter space for a range of input choices.  We leave it for future work to further develop such particle-physics models, potentially including form factors that could reduce the cross section and change the kinematics of the struck nuclei.  We expect that this would shift but not eliminate the regions, e.g., reducing the cross section would allow more DM to reach meteor altitudes.

Figure~\ref{fig:full_exclusion_region_final} also shows that there is much sensitivity to be gained by also taking into account velocity in the data analysis.  If future meteor radar experiments are sensitive to the full DM velocity range, the sensitivity would improve by orders of magnitude, even with no increase in exposure. Since targets with typical DM velocities of a few hundred km/s cannot be meteors (the flux of interstellar meteors is negligible), the backgrounds to such searches would be low.  Here we conservatively use flux alone, because only a small subset of the meteors have well-defined velocities for the CUAM radar system considered here, since the system was built for a wholly different purpose, for which velocity information was not required for every echo. This can likely be improved upon in future work by implementing additional velocity-determining methods to the data.

The sensitivity of meteor-detector searches for DM can be improved far beyond even what we project in Fig.~\ref{fig:full_exclusion_region_final}.  First, larger datasets would extend our sensitivity to larger mass. Second, using data for meteor observations below 70 km altitude would probe smaller DM cross sections. Other meteor-observation techniques (like the photographic probes used by Desert Fireball Network to observe fireballs~\cite{DFNconstraints}) could help.  Third, if the meteor background were better understood, the sensitivity to DM signals would be set by the square root of meteor events in an analysis bin (i.e., the statistical uncertainty) instead of the full number.  Fourth, the DM rate is expected to vary slightly over the course of the year (annual modulation)~\cite{Freese:2012xd}. The meteor backgrounds are also modulated, but differently (see Ref.~\cite{meteorScienceBook}, page 114), which could be used to improve sensitivity.

This new radar-based technique for probing DM is important for several reasons.  First, it is independent of and complementary to other techniques.  Second, it may help probe some of the remaining open regions as well as regions for which the robustness of prior constraints may be doubted.  Third, it provides differential sensitivity to the DM mass and cross section, which is the best way to follow up any hints found by other techniques.  Although we have focused on simple DM candidates, our calculations could be extended to cover more exotic new-physics candidates (charged DM, strangelets, primordial black holes, etc), taking advantage of Earth's atmosphere as the largest conceivable cloud-chamber detector.


\begin{acknowledgments}

We thank the anonymous referee for their helpful comments. We are also grateful for helpful discussions with Javier Acevedo, Christopher Allen, Dave Besson, Joseph Bramanate, Aaron Vincent, and especially Sebastian Ellis, Johan Kero, Annika Peter, and Juri Smirnov. 

C.V.C. and J.F.B. were supported by NSF Grant No.\ PHY-2012955. C.V.C. was also supported by the Arthur B. McDonald Canadian Astroparticle Physics Research Institute.  S.P. was supported by NSF Grant No.\ PHY-2012980. The McMurdo meteor radar is supported by NSF Grant No.\ OPP-1543446. Research at Perimeter Institute is supported by the Government of Canada through the Department of Innovation, Science, and Economic Development, and by the Province of Ontario.

\end{acknowledgments}


\clearpage
\appendix
\counterwithin{figure}{section}

\centerline{\Large {\bf Appendices}}


\section{Atmospheric Density Model}
\label{App:AtmModels}

\begin{figure}[b]
    \centering
    \includegraphics[width=\columnwidth]{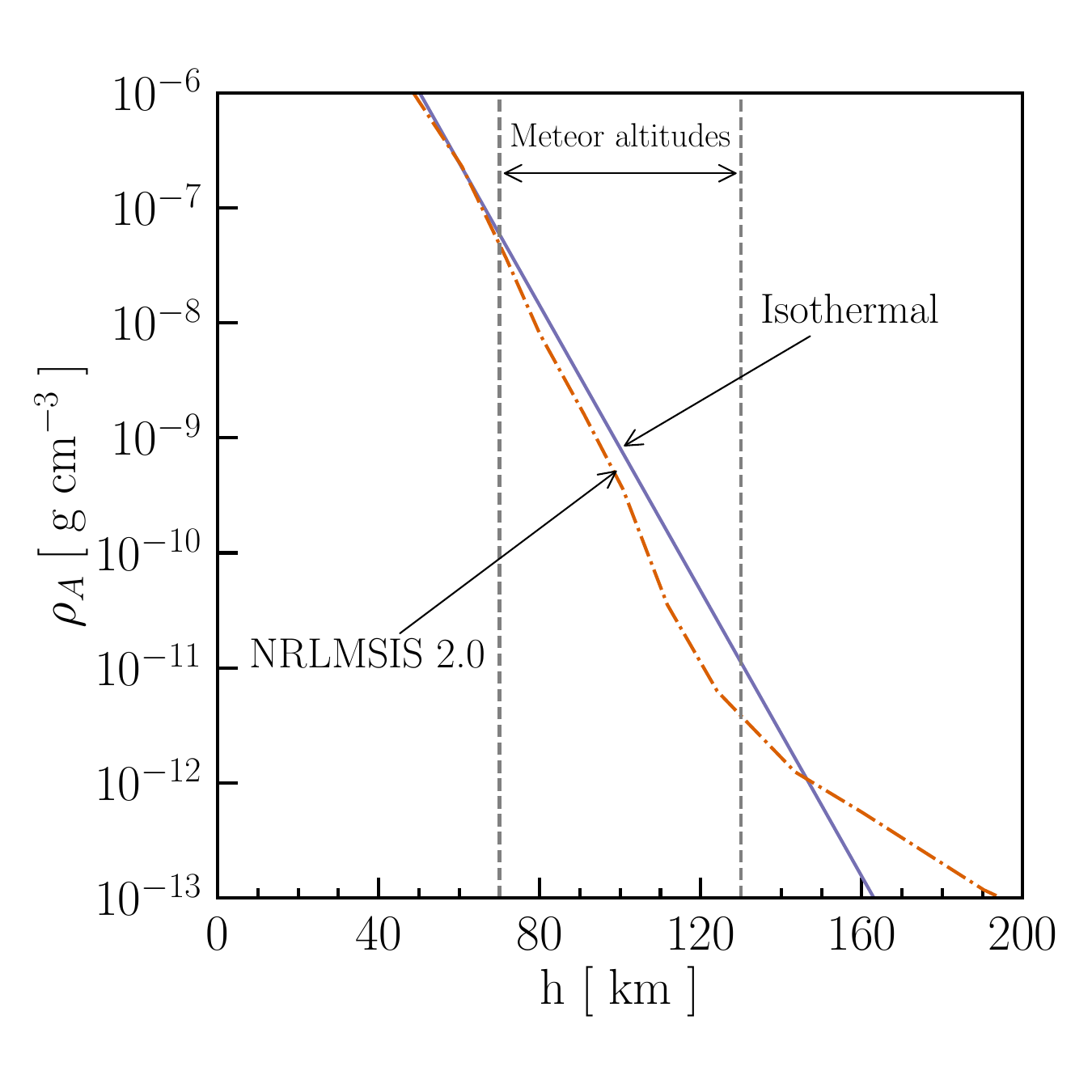}
    \caption{Atmospheric mass density as a function of altitude, comparing the simple isothermal model~\cite{atmModelsLink} and an empirical model~\cite{latestATMModel} (with nitrogen only).}
    \label{fig:comparingAtmModels} 
\end{figure}

Figure~\ref{fig:comparingAtmModels} shows that the isothermal model we use, i.e., a density profile given by Eq.~(\ref{eq:isothermalatm}) with a scale height $H=7$ km, is a good approximation for our precision goals within the meteoric altitude range of 70--130 km.

\section{Meteor and DM Velocity Distributions}
\label{App:VelocityDistributions}

Meteors are typically detected at altitudes of 70--130 km with entry velocities (at 130 km) of approximately 11--70 $\textrm{ km/s}$, corresponding to the escape speed from Earth and that of the solar system near Earth (taking into account Earth's motion).  Objects with higher velocities (up to $800 \textrm{ km/s}$, the escape speed of the Galaxy, taking into account the solar system's motion), which would be on hyperbolic orbits, may be of interstellar origins.  Their flux is low, and the details are uncertain~\cite{interstellar_meteor_flux}.

\begin{figure*}[t]
    \centering
    \includegraphics[height=\columnwidth]{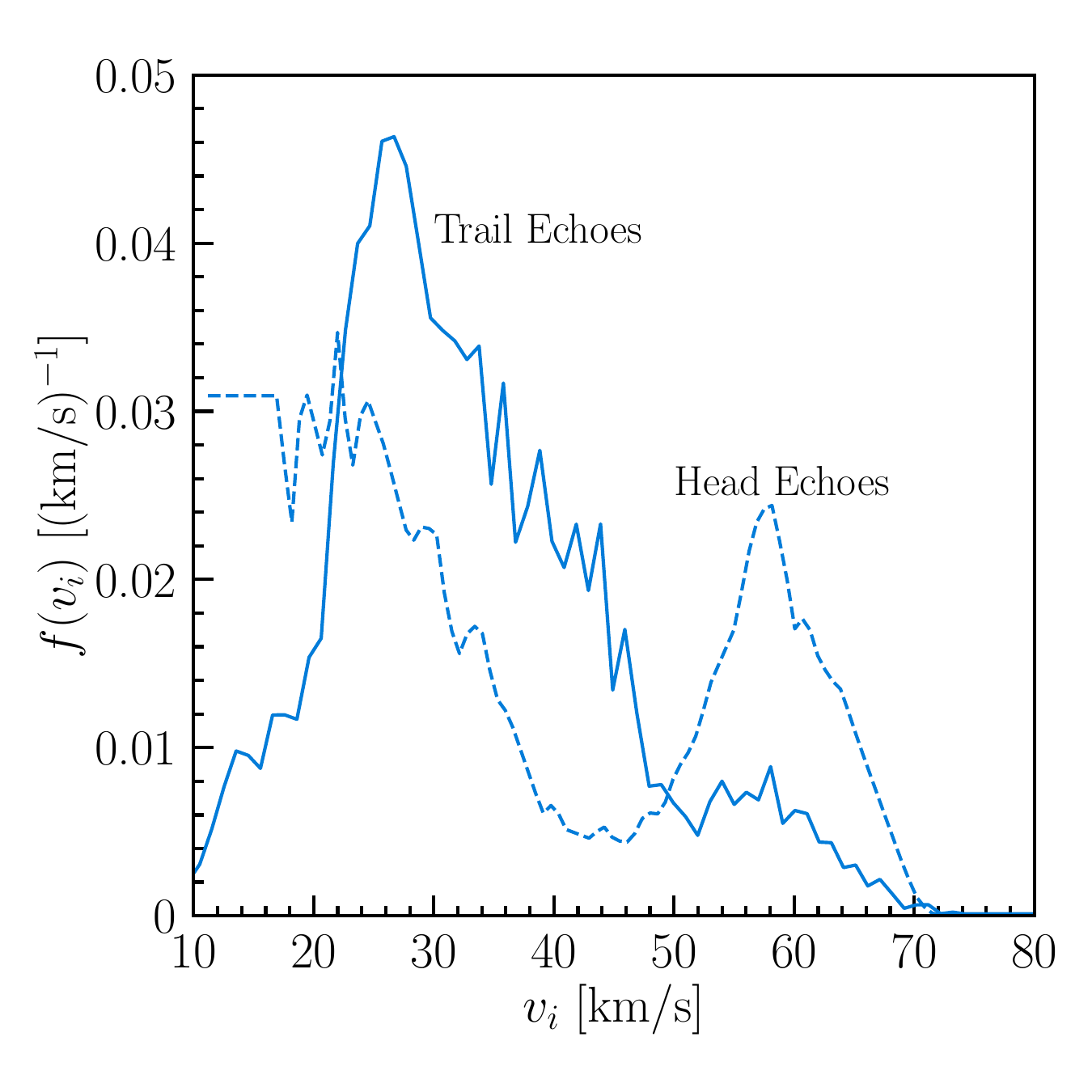}%
    \includegraphics[height=\columnwidth]{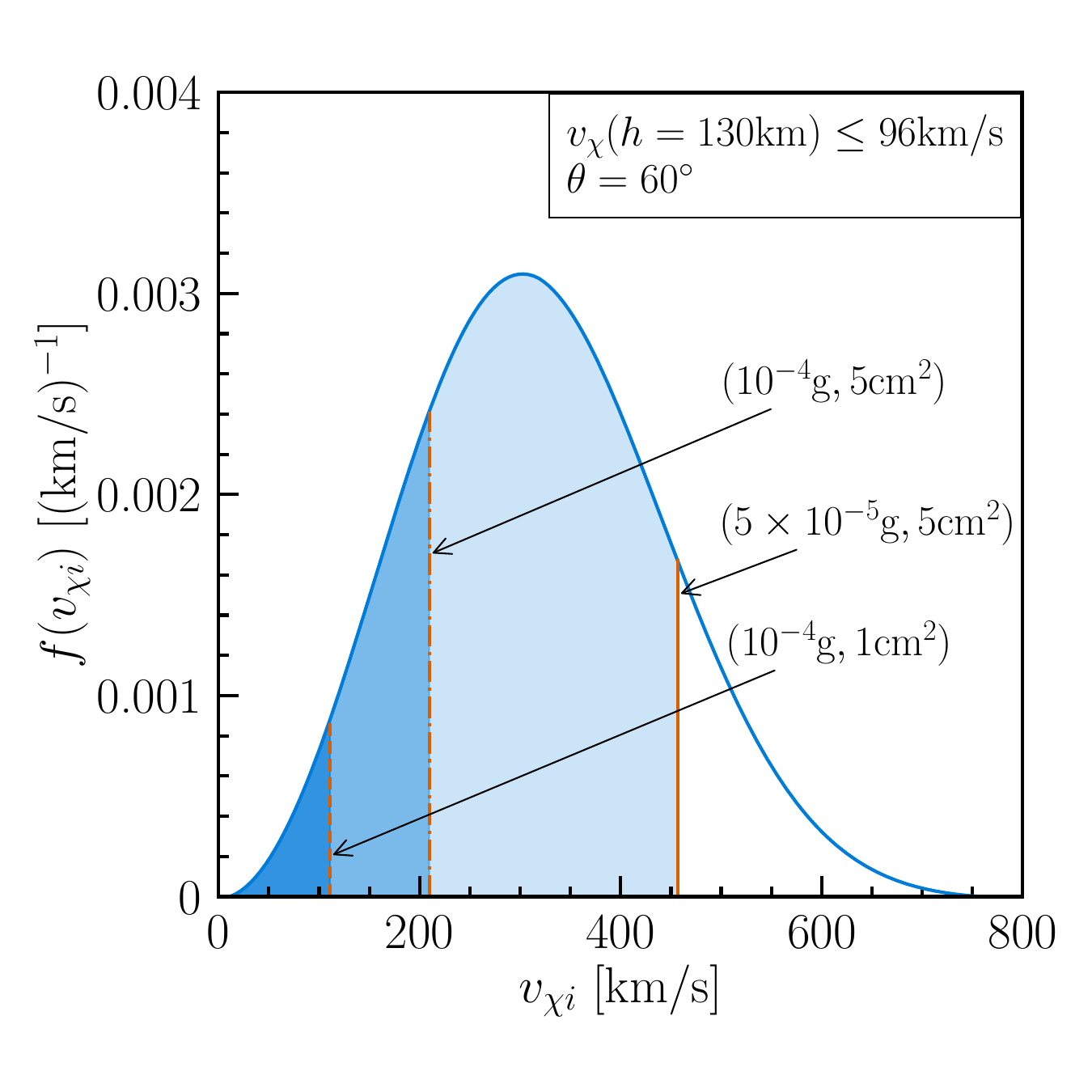}%
    \caption{ 
    {\bf Left:} Observed normalized initial-velocity distributions at 130 km altitude for head echoes (SMUR~\cite{MUpaper}) and trail echoes (CUAM~\cite{AMradarlink}). Note that the trail-echo distribution is based on the $\sim 5\%$ of events with well-constructed velocities.   
    {\bf Right:} The full DM velocity distribution at the top of the atmosphere is shown in solid blue. The shaded regions are the portions of this distribution that would be slowed to below 96 km/s at an altitude of 130 km, for a $60\degree$ zenith angle and several DM mass-cross section pairs. 
    } 
    \label{fig:dm_and_meteors_vel_dist}
\end{figure*}

Figure~\ref{fig:dm_and_meteors_vel_dist} compares the DM velocity distribution $f(v_{\chi i})$ at the top of the atmosphere with the initial-velocity distributions (at 130 km altitude) of sporadic meteors from SMUR~\cite{MUexpSettings} and CUAM~\cite{AMradartypeInterferometry}, which are designed for solar-system meteors.  We calculated the latter from a week's worth of raw data. The DM velocity distribution is based on the Standard Halo Model~\cite{Evans:2018bqy} and includes the effect of Earth's gravity at low velocities. Also shown are the maximum entry velocity at the top of the atmosphere such that DM velocity at 130 km altitude is within radar sensitivity. For SMUR, we use the range 11--96 $\textrm{ km/s}$, while for CUAM, we use 11--70 $\textrm{ km/s}$.  For projected sensitivities, we remove this restriction and use the full velocity range of  11--800 $\textrm{ km/s}$, assuming that future experiments could detect very fast-moving meteors.

\section{Exclusion Regions}
\label{App:ExcRegDetails}

\begin{figure*}[t]
    \centering
    \includegraphics[height=\columnwidth]{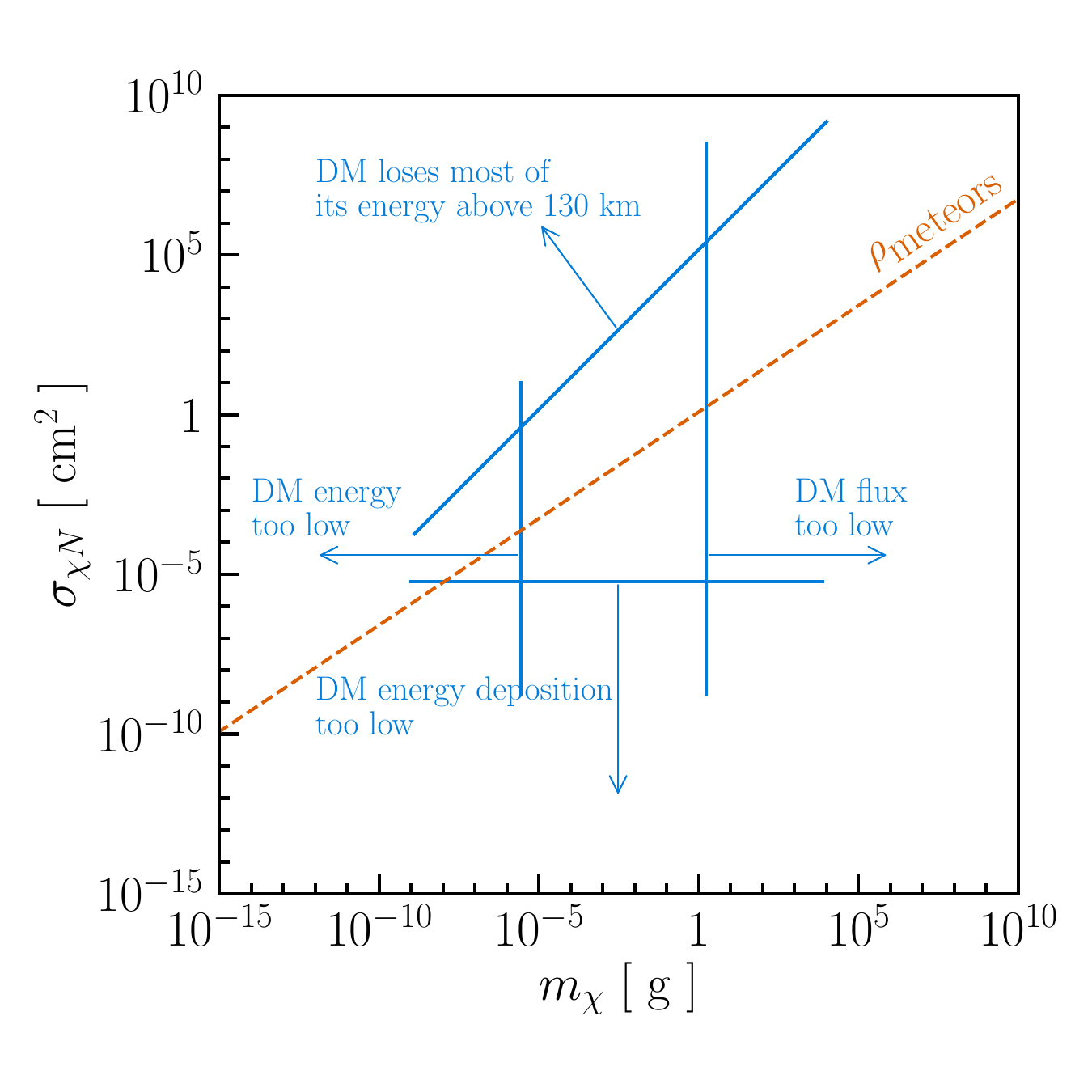}%
    \includegraphics[height=\columnwidth]{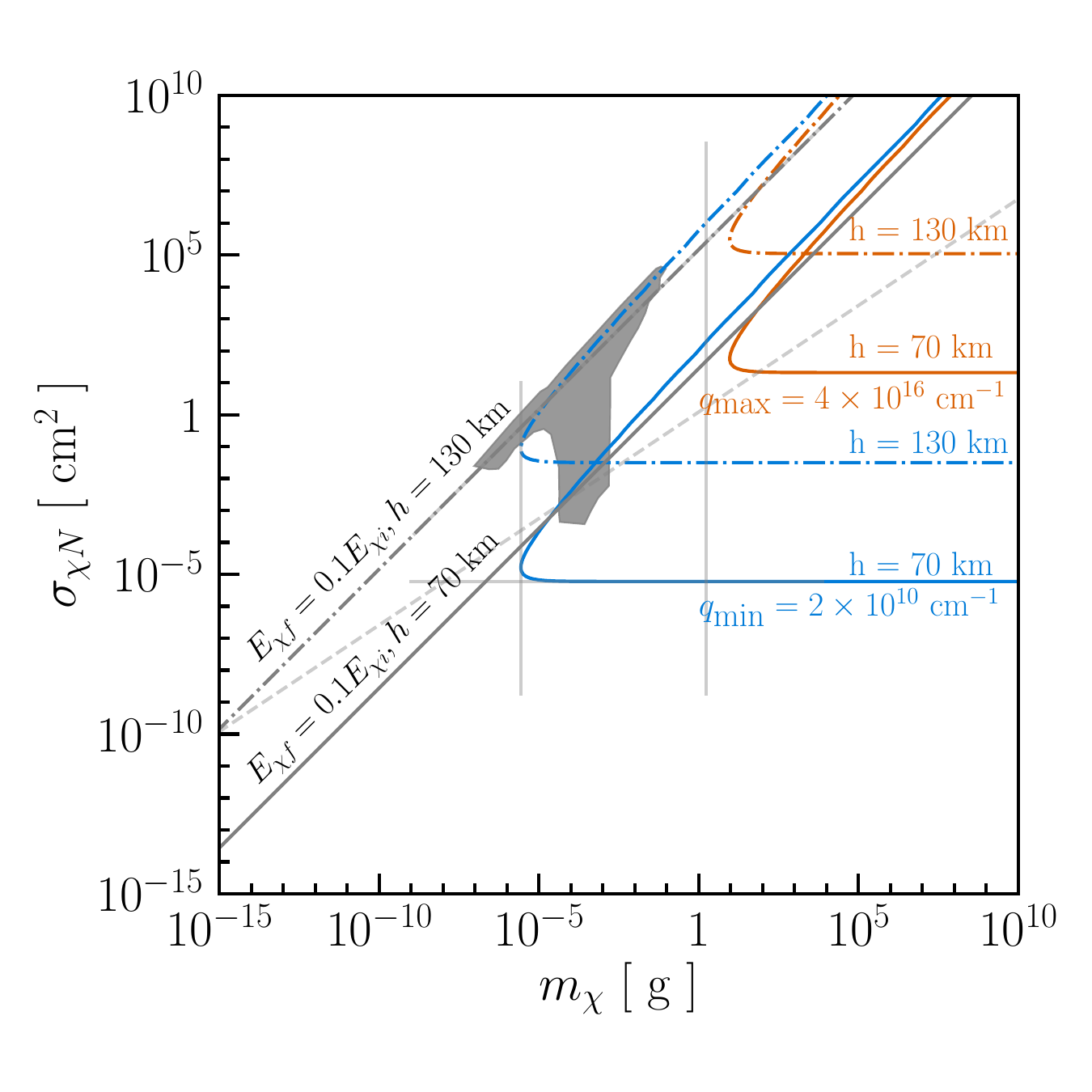}%
    \caption{Estimated constraints in the DM mass and cross section plane compared against the actual constraints for head echoes detectable by the SMUR radar.  
     {\bf Left:} Estimates of the boundaries (blue solid) for head echo constraints. Also shown is a line (orange dashed) corresponding to sporadic meteors with average density of $\sim 1 \textrm{ g cm}^{-3}$. {\bf Right:} Head-echo constraints (gray filled region) overlaid on top of the estimated region. Also shown are lines corresponding to DM candidates that lose $90\%$ of their initial energy at 130 km (gray dash-dotted) and 70 km (gray solid) altitudes, and contours corresponding to the minimum/maximum detectable line density for the SMUR radar, or equivalently the minimum/maximum detectable RCS, (blue/orange) at 130 km (dash-dotted) and 70 km (solid) altitudes.
    } 
    \label{fig:estimatesHeadEchoes}
\end{figure*}

Fig.~\ref{fig:estimatesHeadEchoes} describes the constraints for head echoes detectable by the SMUR radar, similar to Fig.~\ref{fig:estimatesTrailEchoes} for trail echoes. 

%


\cleardoublepage
\bibliography{main}


\end{document}